\documentclass[preprint,12pt,authoryear]{elsarticle}

\usepackage{amssymb}

\usepackage{float}
\usepackage{amsmath}
\usepackage{xcolor}
\usepackage{placeins}
\usepackage{hyperref}
\usepackage{multirow}
\usepackage{eurosym}

\DeclareMathOperator*{\argmin}{\arg\!\min}

\emergencystretch=\maxdimen
\journal{}

\begin{document}

\begin{frontmatter}



\title{An adaptive standardisation methodology for Day-Ahead electricity price forecasting}


\author[label1,label2]{Carlos Sebastián}

\affiliation[label1]{organization={Fortia Energía},
             addressline={Calle de Gregorio Benítez},
             city={Madrid},
             postcode={28043},
             country={Spain}}

\affiliation[label2]{organization={Universidad Politécnica de Madrid},
            city={Madrid},
            country={Spain}}
\ead{carlos.sebastian@alumnos.upm.es}
            
\author[label3,label4]{Carlos E. González-Guillén}

\affiliation[label3]{organization={Departamento de Matemática Aplicada a la Ingeniería Industrial, Escuela Técnica Superior de Ingenieros Industriales, Universidad Politécnica de Madrid},
            addressline={Calle de José Gutiérrez Abascal}, 
            city={Madrid},
            postcode={28006},
            country={Spain}}

\affiliation[label4]{organization={Instituto de Ciencias Matemáticas (CSIC-UAM-UC3M-UCM)},
            addressline={Calle Nicolás Cabrera}, 
            city={Madrid},
            postcode={28049},
            country={Spain}}
            
\author[label5]{Jesús Juan}

\affiliation[label5]{organization={Laboratorio de Estadística, Escuela Técnica Superior de Ingenieros Industriales, Universidad Politécnica de Madrid},
            addressline={Calle de José Gutiérrez Abascal}, 
            city={Madrid},
            postcode={28006},
            country={Spain}}

\begin{abstract}
The study of Day-Ahead prices in the electricity market is one of the most popular problems in time series forecasting. Previous research has focused on employing increasingly complex learning algorithms to capture the sophisticated dynamics of the market. However, there is a threshold where increased complexity fails to yield substantial improvements. In this work, we propose an alternative approach by introducing an adaptive standardisation to mitigate the effects of dataset shifts that commonly occur in the market. By doing so, learning algorithms can prioritize uncovering the true relationship between the target variable and the explanatory variables. We investigate five distinct markets, including two novel datasets, previously unexplored in the literature. These datasets provide a more realistic representation of the current market context, that conventional datasets do not show. The results demonstrate a significant improvement across all five markets using the widely accepted learning algorithms in the literature (LEAR and DNN). In particular, the combination of the proposed methodology with the methodology previously presented in the literature obtains the best results. This significant advancement unveils new lines of research in this field, highlighting the potential of adaptive transformations in enhancing the performance of forecasting models.
\end{abstract}

\begin{highlights}
\item A new methodology for Day-Ahead electricity price forecasting is proposed.
\item The focus of the methodology is to perform an adaptive standardisation of the series to be predicted, making all the time points of the series comparable to each other.
\item Two new current data sets (2019-2023) are made public due to the lack of studies in the literature related to the current electricity market context. 
\item The proposed methodology improves the results of the state-of-the-art, with statistical evidence, in four different markets and in two periods characterised by completely different behaviour, highlighting the robustness to different scenarios.
\item The best results are obtained by combining models. In particular, these are obtained by combining the proposed methodology with existing approaches from the literature.
\end{highlights}


\end{frontmatter}


\section{Introduction}
\label{sec:introduction}

\begin{figure}[h]
\centering
\includegraphics[width=\textwidth]{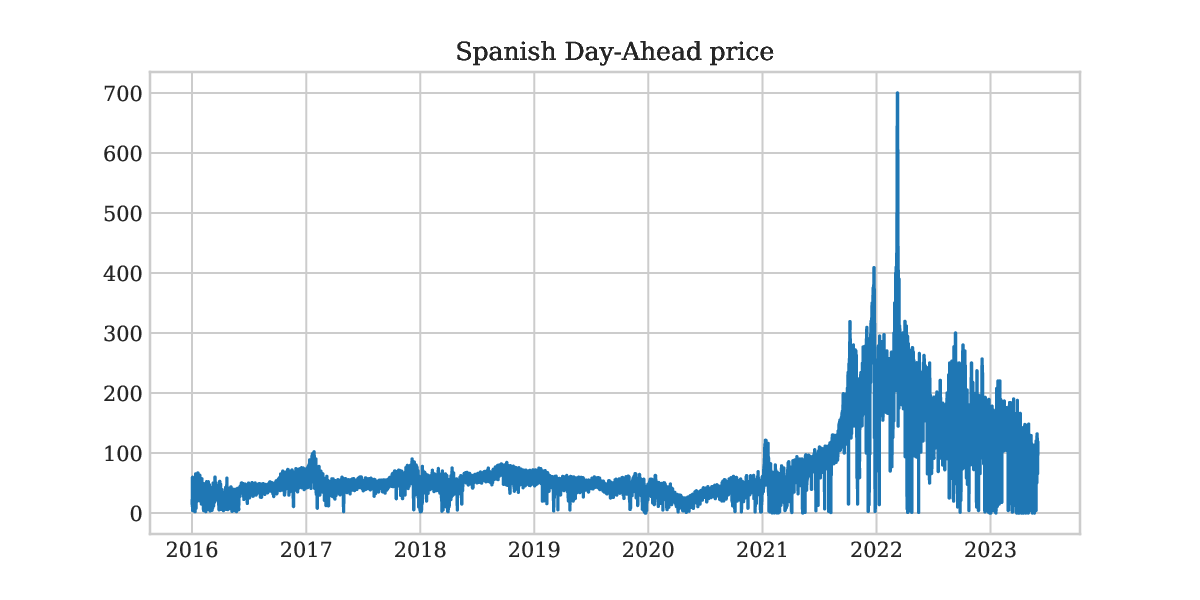}
\caption{Spanish Day-Ahead market prices through the years}
\label{fig:spanish_day_ahead}
\end{figure}

Electricity is one of the most vital commodities in today's society. However, operating the electricity system presents unique challenges that must be addressed such as the balance between production and consumption, the independent hourly or quarter-hourly scheduling of different markets, the technical constraints of the generation program related to the security of the network, etc. In addition, significant changes in its behaviour occur: the increasing presence of renewable energies, the dynamics of external prices that affect generation costs and the impact of socio-economical factors are some examples. These factors contribute to the constant evolution and considerable uncertainty in the market, particularly  concerning price predictions. We illustrate this phenomenon using the Spanish Day-Ahead market, although similar behaviors can be observed in other European countries (Figure \ref{fig:spanish_day_ahead}).\\

From 2016 to 2023, a noticeable shift in price dynamics emerge towards the end of 2021. As a result, we can observe three distinct phases: a period of stability, a subsequent phase characterized by increased volatility, and an intermediate transitory interval.\\

While the electricity market has been gaining attention over the years \citep{hong2020energy}, and a rich literature related to the Day-Ahead market price forecasting has been developed \citep{lago2021forecasting}, most studies focus on older stable periods that do not reflect the peculiarities of the current market.\\

This paper contributes to the existing literature as follows:

\begin{enumerate}
\item We propose a new methodology for Day-Ahead electricity price forecasting (EPF) that is characterized by ensuring comparability among all periods of the time series through an adaptive standardisation approach. This attribute makes the predictive model under consideration flexible, allowing it to effectively respond to evolving situations and improve its performance accordingly.

\item We are providing access to two new datasets that span from 1st January 2019 to 31st May 2023. These datasets are made available not only for the purpose of replicating the results, but also to evaluate new models, enabling them to accurately reflect and represent the conditions of the current market.

\item An improvement is provided from a statistical perspective rather than from a machine learning perspective, which has been the main focus of research in recent years. This progress represents a significant development that has not been observed since \cite{uniejewski2016automated}.
\end{enumerate}

The remainder of the paper is organized as follows: Section \ref{sec:previous_work} performs a review of state-of-the-art trends and models in Day-Ahead EPF. Section \ref{sec:new_model} presents the novel methodology proposed and the rationale behind it. Finally, Section \ref{sec:results} tests the models and discusses the results, ending with the conclusions in Section \ref{sec:conclusions}.

\section{Previous work}\label{sec:previous_work}

\subsection{Electricity price forecasting (EPF)}

EPF is an open field in which a wide variety of tasks are included, mainly depending on the market being dealt with: Day-Ahead market, Intra-Day markets or Balancing markets. Among these, the Day-Ahead market has garnered the most significant attention. While the regulatory framework of this market varies across countries, its structure, follows a standard pattern in Europe: on day D, prior to a designated hour H, all market participants are required to submit their bids for purchasing or selling energy for each hour of the subsequent day, D+1. The price of each hourly period is established independently through an auction-based format.\\

As explained in \cite{ziel2018day}, mainly two approaches can be considered in EPF and, indeed, in other forecasting problems of the electricity market:

\begin{enumerate}
\item Consider a high-frequency univariate time series, typically observed at an hourly interval, where a single global model is used for analysis and prediction.

\item Consider a multivariate time series, where the subsequent 24 hours must be predicted using a single multivariate model. This approach can be transformed in an univariate perspective by treating each hour independently and, therefore, using separate models for each hour.
\end{enumerate}

The first option is the standard approach when working with machine learning models, as they typically require vast amounts of data. In contrast, the second option with 24 separate models, is commonly found in statistical models. However, according to \cite{ziel2018day}, neither approach dominates the other. It is the combination of both methods that yields superior results in terms of predictive performance.\\

In a distinct research direction, it has been studied the application of Functional Data Analysis (FDA) within the electricity market, as can be seen in \cite{chen2021review}. FDA treats individual time series as discrete observations derived from a continuous function. Building upon this concept, two relevant studies focusing on price prediction are those by \cite{chen2017adaptive} and \cite{jan2022short}, both of which showcase promising outcomes and advancements in this domain.\\

Nevertheless, establishing state-of-the-art models in the field of EPF, particularly within the Day-Ahead market, presents a complex challenge due to a multitude of factors highlighted by \cite{lago2021forecasting}:

\begin{itemize}
\item It is typical for different studies to come to contradictory conclusions, especially when comparing classical statistical methods with machine learning techniques.

\item Similarly, comparisons are often drawn with basic models that fall considerably short of achieving the performance levels exhibited by the best known models. Furthermore, studies often analyze datasets that have not been previously explored or documented in the existing literature.

\item The test period is often relatively short, which introduces the possibility of overlooking special situations and failing to adequately assess the annual seasonality inherent in price dynamics. As a consequence, the obtained results may be influenced by the specific choice of the test window, potentially limiting their generalizability.

\item Sometimes the results cannot be replicated, not allowing the results to be tested by other community members.

\end{itemize}

As we strongly agree with these statements, we align ourselves with the definition of state-of-the-art models as presented in the review by \cite{lago2021forecasting}.\\

The Lasso Estimated AutoRegressive (LEAR), proposed in \cite{uniejewski2016automated}, is considered as the best statistically based model. It is is an autoregressive model with exogenous variables (ARX) to which an L1 regularisation is applied.\\

The DNN proposed in \cite{lago2018forecasting} is considered one of the best machine learning based models. It has a simple structure consisting of a fully connected neural network with two hidden layers and an output layer comprising 24 neurons, each corresponding to a specific hour. So, instead of having one model for each hour as in the previous case, there is just one model for the whole series.\\

From a machine learning perspective, a new model has achieved better results: NBEATSx \citep{olivares2023neural}, an extension of the NBEATS model \citep{oreshkin2019n} that allows for the inclusion of exogenous variables. One advantage of NBEATSx is the ability to perform an analysis to understand how each potential source of information affects the predictions, which is not commonly found in models of this nature. When applied to typical datasets in the literature, the results obtained by this model are superior to those presented by the previously mentioned neural network. However, it is worth mentioning that the improvement achieved by NBEATSx may not always reach statistical significance in certain markets according to the Giacomini-White test \citep{giacomini2006tests} carried out by the authors. Another aspect to consider is the increase in computational complexity associated with this model.\\

The X-Model presented in \cite{ziel2016electricity} is also noteworthy, as it follows a completely different approach to the usual one. Instead of the classical time series modelling, it tries to model the supply and demand curves that determine the price for each hourly period. The results are quite positive, even capturing non-linear behaviour.\\

In the EPF context, it has been acknowledged that the behavior of the target series exhibits temporal variations. To address this issue, \cite{nasiadka2022calibration} propose the integration of change point methods, aiming to identify historical windows in the past that align with the current market conditions. These selected windows are then incorporated into the training set. However, it is important to recognize that these change point methods may not always accurately identify the relevant windows due to the complexity of the underlying market dynamics. The effectiveness of such methods relies on the assumption that the identified change points truly reflect the shifts in the market behavior. In practice, there can be instances where the change point detection may fail to capture the subtle nuances or abrupt changes in the series, leading to potential inaccuracies in the selection of training data. Moreover, a significant portion of the available information in the historical data remains unused. Thus, while change point methods offer a valuable approach, their limitations should be considered.\\

Although probabilistic forecasting is beyond the scope of this study, it is worth noting that a significant portion of the EPF field is devoted to this area. Therefore, it is necessary to mention some of the main works within this particular trend. A satisfactory idea was introduced in \cite{nowotarski2015computing}: applying quantile regression \citep{koenker1978regression} using the predictions obtained by point estimators as explanatory variables (QRA). Given the good results of the LEAR model in point estimation, \cite{uniejewski2021regularized} propose the use of quantile regression following the same philosophy, but applying L1 regularisation on the loss function, so that an automatic selection of variables is performed, improving the results. Complex recurrent or convolutional neural network structures in a probabilistic context are compared in \cite{mashlakov2021assessing} but focusing on the electricity market in general, not only on price. More classical techniques such as bootstrap over residuals \citep{efron1992bootstrap} to obtain probabilistic results have also been used for price forecasting in the electricity market in \cite{l2021optimal}. A more in-depth study on the price appears in \cite{marcjasz2022distributional} through the use of distributional neural networks, used for the first time in this field. The results are better than those obtained by applying QRA to the LEAR model or to the neural network of \cite{lago2018forecasting}, although not entirely satisfactory as observed by the number of hours that pass the Kupiec test \citep{kupiec1995techniques} at 50\% and 90\%. It is also worth mentioning that the use of the novel framework of conformal prediction \citep{vovk2005algorithmic} has also been applied in EPF \citep{kath2021conformal}, where it is concluded that valid prediction intervals can be obtained on the predictions made, improving on several metrics of QRA.

\subsection{Adaptive transformation schemes}
When referring to adaptive transformation methods, we specifically mean time series transformations that utilize a window of past observations rather than the entire historical data to achieve stationarity. It is important to note that we do not consider the use of rolling windows alone as a form of adaptive transformation because the models that employ rolling windows typically utilize data solely from within the window to process information and make predictions.\\

The first adaptive transformation technique was introduced in \cite{ogasawara2010adaptive}. In this case, a modified version of the conventional min-max normalisation method is proposed to address specific challenges associated with the traditional approach, particularly those related to volatility. The technique is thoroughly analyzed across various non-stationary series, consistently showing better results.\\

In \cite{passalis2019deep}, a deep learning approach is employed. The proposed method, known as Dynamic Adaptive Input Normalization (DAIN), includes the data transformation as trainable parameters within the neural network architecture. This allows the optimal transformation to be learned and automatically adjusted during the training process, similar to other network parameters. In this way, each time the network is trained, it is calibrated to the current context of the system that is being predicted. Empirically, the results demonstrate an improvement when compared to conventional approaches.\\

Lastly, \cite{gamakumara2023conditional} propose an estimation technique for the conditional mean and conditional variance using Generalized Additive Models (GAM), \citep{hastie1990generalized}. By employing GAM models, the estimation process allows the conditional standardisation of the data by effectively removing sources of variation external to the time series itself. Although this is not an adaptive transformation as such, this approach proves to be more suitable for non-stationary time series compared to classical methods, enabling improved modeling and analysis of the underlying dynamics.

\section{Proposed methodology}\label{sec:new_model}
A widely adopted methodology, regardless of whether statistical or machine learning-based methods are employed, is to transform non-stationary price series into a more stationary form before implementing the learning algorithm. This involves applying one of several available transformations that aim to achieve a constant mean and stabilize the variance \citep{uniejewski2017variance}.These transformations are mainly based on some form of standardisation of  the data and are justified for several reasons. In the context of neural networks, which are commonly used in Electricity Price Forecasting (EPF) problems, employing standardised features enhances the stability of the numerical algorithms that solve the underlying optimisation problems. When working with linear models, it has been observed (see Section \ref{sec:previous_work}) that using some kind of regulariser improves the results. If the variables are not on the same scale, this technique becomes less effective, as the penalty imposed by the regularisation does not uniformly affect all variables. Regardless of the transformation used, they are computed in the training set for subsequent application in the validation and test sets.  However, it is important to acknowledge that this methodology assumes the absence of any change in the joint distribution of the data, implying no occurrence of dataset shift \citep{quinonero2008dataset}. Nonetheless, a quick examination of Figure \ref{fig:spanish_day_ahead} shows the presence of potential shifts within the data.\\

The models presented in Section \ref{sec:previous_work} implicitly assume that the series is stationary in both mean and variance. There have also been studies that assume that the variance of the process can vary over time, for example, by considering GARCH-type models. However, in \cite{janczura2023arx}, it is shown that considering this fact alone does not produce improvements from the perspective of point forecasting. We believe that a further step should be taken in this direction in order to model prices correctly.\\

Looking at the behaviour of prices, it is reasonable to consider that consecutive intervals of varying length in the time series exhibit stationarity, or at least a constant mean and variance. This means that the time series is piecewise stationary in mean and variance. Therefore, in order to perform proper modelling, the change points in the behaviour of the series should be correctly detected and each stability period should be transformed using the correct estimation of the mean and standard deviation of each period.\\

In view of this fact, let $p_d^h$ be the price series corresponding to the hour $h$ in the day $d$, the following model is proposed\begin{equation}\label{eq:model}
\begin{cases}
p_d^h &= \mu(X_d) + \sigma(X_d)u_d^h \equiv \mu_d + \sigma_d u_d^h\\
u_{d+1}^h &= f(X_d, u_d^h, u_{d-1}^h, \dots) + \varepsilon_{d+1}^h, \; \mathbb{E}\left[\varepsilon_{d+1}^h\right] = 0
\end{cases}
\end{equation}

We propose the following parameter estimation methodology:
\begin{equation}\label{eq:params}
\begin{cases}
\hat{\mu}_{d} &= \dfrac{1}{24v} \displaystyle\sum_{k =d-v-1}^{d-1}\sum_{h=1}^{24}p_k^h, \; \text{for each } d\\
\hat{\sigma}_{d} &= \sqrt{\dfrac{1}{24v} \displaystyle\sum_{k = d-v-1}^{d-1}\sum_{h=1}^{24}\left(p_k^h - \hat{\mu_{d}}\right)^2}, \; \text{for each } d \\
\end{cases}
\end{equation}
where $v$ is the lenght of a rolling window in days, $f$ is a learning algorithm and $X_d$ is the set of explanatory features available before the end of the biding period for day $d$.\\ 

Although this approach to parameter estimation does not guarantee that the variance of \(\epsilon_{h}^{d}\) remains constant across all \(d\) and \(h\), it operates under the assumption that the series is piecewise stationary in terms of variance. Deviations from this assumption typically occur near behavioral change points, which are inherently unpredictable. If the chosen window \(v\) is sufficiently small and entirely within a stable period, the model effectively estimates the mean and variance for that period using the latest \(24 \times v\) data points. However, if the subsequent observation corresponds to a change point, the standardisation process will be incorrect, introducing errors proportional to the magnitude of that change. Similarly, if the window spans multiple stability periods, the parameters are likely to be misestimated. The extent of this misestimation depends on the proportion of data from the newer period and the intensity of the change in behavior. Consequently, if the forthcoming data point to be predicted represents a change point or belongs to a very recent regime, predicting its value accurately becomes challenging.\\

Note that the terminology used in (\ref{eq:model}) aligns with the proposal of \cite{gamakumara2023conditional} where both the mean and standard deviation are modelled as a function of a number of explanatory features. In our case, the effects of external variables are directly incorporated through the model $f$, with the notable exception of outlier accommodation, which is detailed below.\\

An important detail in the estimation of parameters (\ref{eq:params}) is that it can be affected by outliers. Thus, instead of working with the original price series $p_d^h$ it is proposed to work with one where the outliers have been filtered out and replaced appropriately. That is, at the time of predicting the observation $p_{d+1}^h$, the price series in the training set will be the one given by $$\tilde{p}_{d'}^h = 
\begin{cases} 
    p_{d'}^h &  \text{if } \,\hat{\mu_{d'}} - \kappa \cdot \hat{\sigma_{d'}} \leq p_{d'}^h \leq  \hat{\mu_{d'}} + \kappa \cdot \hat{\sigma_{d'}}  \\
    \underset{\substack{t\in \{ 1, \dots, v\} \\ h'\in \{1, \dots, 24\}}} {\text{Median}} \left\{p_{d'-t}^{h'} \right\}  & \text{in other case}
\end{cases},
$$ where $\hat{\mu_{d'}}$ and  $\hat{\sigma_{d'}}$ are computed as in (\ref{eq:params}) for a window of $v$ days.\\

\begin{figure}[htp]
\centering
\includegraphics[width=\textwidth]{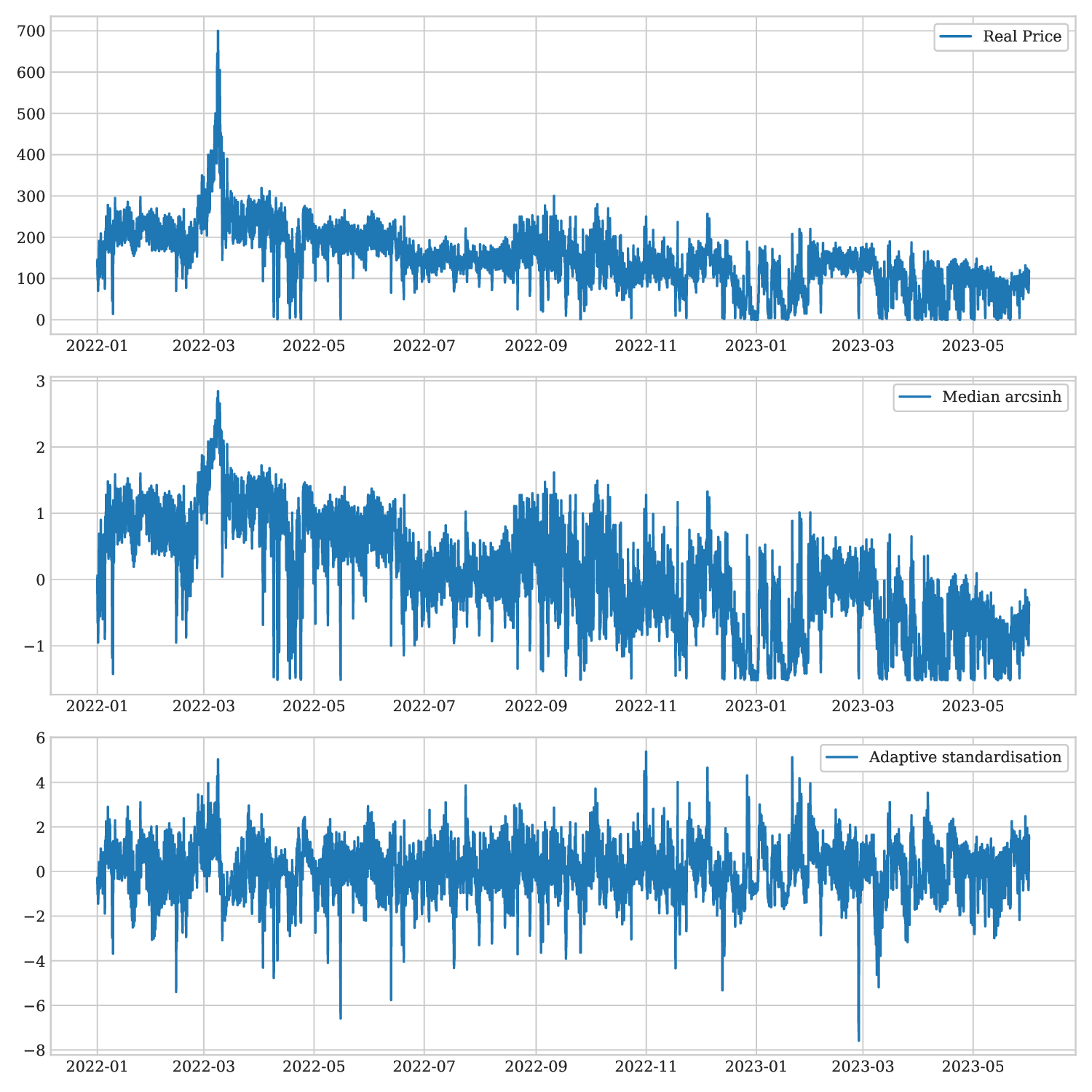}
\caption{Comparison of the current price in the Spanish Day-Ahead market during hour 20 (top), its median-arsinh transformation (middle) and the resulting process derived of applying our methodology with $v=7$ days, 168 hours (bottom)}
\label{fig:standarization_schemes}
\end{figure}

Figure \ref{fig:standarization_schemes} presents a comparison between the current price series in the Spanish Day-Ahead market and the series obtained using a median-arcsinh standardisation scheme, which has been applied in the EPF literature \citep{uniejewski2018efficient}, and another series representing the $u_d^h$ process resulting from our proposed methodology.\\

It is evident from the figure that our methodology yields a significantly more stationary series, which aligns with the main objective of applying such transformations. The fundamental idea of the transformation is to encapsulate the effects of potential shifts in the $\mu_d$ and $\sigma_d$ parameters, ensuring that the resulting process $u_d^h$ is minimally affected by such situations. Additionally, implementing the adaptive standardisation through a reasonably sized rolling window enables a rapid response to these shifts.\\

As this adaptive standardisation mitigates shifts in the behaviour of the price series, the unmodified explanatory features are sub-optimal candidates for the prediction of transformed prices. They also need to be transformed. Furthermore, this step is crucial to avoid spurious regressions caused by non-stationary explanatory variables \citep{harris2003applied}. We choose to apply the rolling standardisation of length $v$, with the parameters $\mu_{d}$ and $\sigma_{d}$ obtained in (\ref{eq:params}) to each one of the variables for non-dummy features and leave dummy features unchanged.\\

\section{Experiments}\label{sec:results}

To evaluate the results, a thorough analysis of the proposed methodology is conducted using the established Python library \texttt{epftoolbox} \citep{lago2021forecasting}. By comparing the use of a conventional transformation scheme in the EPF literature (median-arcsinh transformation, \cite{uniejewski2018efficient}) with the one we propose, we can show the significance of our approach and its impact on the final model. The models are retrained each day, as one should do in a real industrial EPF situation. For the adaptive standardisation, we have selected $v = 7$ days ($168$ hours) due to the seasonality of prices (caused by the electricity load). Although other windows such as 14 days or 21 days could also be viable options, the smallest ``weekly'' window has been selected to allow for quick adaptation to new regimes. Logically, $v$ can be chosen by the user based on the need of the problem to be solved. In fact, the choice of this window could vary between the different datasets considered, with a larger window for more stable series and a smaller window for cases with strong changes in behaviour. However, for simplicity, only one value has been considered.\\

\subsection{Learning algorithms}

We consider the LEAR model and the neural network both presented in Section \ref{sec:previous_work} and available in the \texttt{epftoolbox} package, as potential learning algorithms $f$ (Equation \ref{eq:model})

\subsubsection{LEAR}

The model specification is

\begin{equation*}
\begin{aligned}
u_d^h = &\sum_{i=1}^{24}\theta_{h, i} \cdot u_{d-1}^i + \sum_{i=1}^{24}\theta_{h, 24 + i} \cdot u_{d-2}^i + \sum_{i=1}^{24}\theta_{h, 48 + i} \cdot u_{d-3}^i + \sum_{i=1}^{24}\theta_{h, 72 + i} \cdot u_{d-7}^i\\
&+ \sum_{i=1}^{24}\theta_{h, 96 + i} \cdot x_{d, 1}^i + \sum_{i=1}^{24}\theta_{h, 120 + i} \cdot x_{d, 2}^i+ \sum_{i=1}^{24}\theta_{h, 144 + i} \cdot x_{d-1, 1}^i \\
&+ \sum_{i=1}^{24}\theta_{h, 168 + i} \cdot x_{d-1, 2}^i+ \sum_{i=1}^{24}\theta_{h, 192 + i} \cdot x_{d-7, 1}^i + \sum_{i=1}^{24}\theta_{h, 216 + i} \cdot x_{d-7, 2}^i \\
&+ \sum_{j=1}^{7}\theta_{h, 240 + j} \cdot {z}^{j}_{d} + \varepsilon_{d}^h
\end{aligned},
\end{equation*}
where $u_{d}^h$ is the standardised price of day $d$ in hour $h$ ($p_d^h$), $x_{d, 1}^h$ and $x_{d, 2}^h$ are two variables of interest associated with the market on day $d$ in hour $h$, usually related to load, wind forecasting or solar forecasting, and \({z}^j_{d}\) is a binary variable that assumes a value of 1 if day of the week $j$ is the same as day $d$, and 0 otherwise. The values of $x_{d, 1}^h, x_{d, 2}^h$ may require a standardisation similar to that of the price. The $\varepsilon_{d}^h$ simply represents noise. Note that this is a model for each hour.\\

The coefficients of the model $\hat{\theta}_h = (\hat{\theta}_{h, 1}, \dots, \hat{\theta}_{h, 247})$ are calculated as $$\hat{\theta}_h = \argmin_{\theta_h} \sum_{d=1}^{N_d}(u_{d}^h - \hat{u}_{d}^h)^2 + \lambda \sum_{i=1}^{247} \vert \theta_{h, i} \vert$$ with $\hat{u}_{d}^h$ the forecast of day $d$ at hour $h$, $N_d$ the number of days in the training set and $\lambda$ a regularization hyperparameter of the model that can be fitted across a multitude of schemes.\\ 

In \cite{lago2021forecasting} the LARS method is used together with the AIC score to optimise the hyperparameter $\lambda$. Due to updates of the \texttt{Scikit-learn} library \citep{pedregosa2011scikit} the application of this method is not so easy for small calibration windows (where the number of observations is smaller than the number of variables). This is why cross-validation has been adopted as a method to determine this hyperparameter. This increases the computation time significantly, but facilitates the determination of $\lambda$ for small calibration windows. In any case, for calibration windows where this problem is not encountered, experiments with the LARS-AIC scheme have been replicated to verify that the results differ only slightly (see \ref{ap:lars}).\\ 

\subsubsection{DNN}

We also consider the DNN proposed in \cite{lago2018forecasting} as a representative of the machine learning based models. The model parameters are estimated using Adam optimization algorithm \citep{kingma2014adam}. We employ the same set of potential input features as employed in the LEAR model. The selection of specific input features is determined as another hyperparameter of the network. To optimize these hyperparameters, we employ a tree-structured Parzen estimator \citep{bergstra2011algorithms}. Notably, hyperparameters unrelated to the explanatory features encompass the following: the number of neurons per layer, the choice of activation function, the dropout rate, the learning rate, the inter-layer normalisation scheme, preprocessing procedures prior to the input layer, weight initialization strategies, and the coefficient associated with the applied L1 regularization.\\

\subsection{Datasets}

\begin{figure}[h]
\centering
\includegraphics[width=\textwidth]{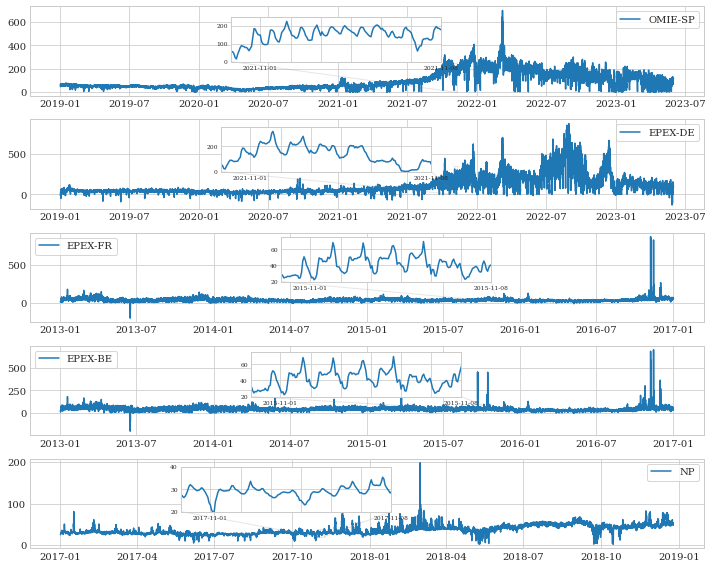}
\caption{Markets considered. Observe the differences between the variability of current markets and previous markets.}
\label{fig:market_prices}
\end{figure}

In this paper, we analyze five datasets, two of which are novel contributions to the literature. These datasets are in the public domain and we make them available to the research community\footnote{\url{https://github.com/CCaribe9/AdaptStdEPF}}, as they offer valuable opportunities for studying and addressing market conditions that are more representative of the current state. The five markets to consider are (Figure \ref{fig:market_prices}):

\begin{itemize}

\item The OMIE-SP market, representing the Spanish electricity market. This market is not available in the Python package. Price data spans from  January 1st, 2019 to May 31st, 2023. Two explanatory variables, namely the day-ahead load forecast and the forecast of renewable energy generation (solar photovoltaic, solar thermal and wind). The data has been obtained through the ESIOS platform\footnote{\url{https://www.esios.ree.es/}}.

\item The EPEX-DE market, representing the German electricity market, is another dataset considered in this study. Data from January 1st, 2019 to May 31st, 2023 has been obtained from the ENTSO-E transparency platform\footnote{\url{https://transparency.entsoe.eu/}}. Similar to the previous market, the same exogenous variables have been considered. However, in this case, the renewable generation forecast does not include the solar thermal generation forecast.

\item The EPEX-FR market, which is the Day-Ahead electricity market in France. The dataset also encompasses the period from January 9th, 2011 to December 31st, 2016. It incorporates two exogenous variables: the day-ahead load and generation forecasts from France. The dataset is accessible from the \texttt{epftoolbox} package.

\item The EPEX-BE market, which is the Day-Ahead electricity market in Belgium. The dataset encompasses the period from January 9th, 2011 to December 31st, 2016. It incorporates the same two exogenous variables from France, which, although surprising, are two of the best regressors for this market \citep{lago2018forecasting}. The dataset is accessible from the \texttt{epftoolbox} package.

\item The NP market, which is the European power market of the Nordic countries. Data spans from January 1st, 2013 to December 14th, 2018. The explanatory features considered are the day-ahead load forecast and the day-ahead wind generation forecast. The dataset is accessible from the \texttt{epftoolbox} package.

\end{itemize}

The test period of every market is detailed in Table \ref{table:test_periods}.\\

\begin{table}[h]
\centering
\begin{tabular}{c|c}
\textbf{Market}  & \textbf{Test period}             \\ \hline
OMIE-SP & 01.01.2022 - 31.05.2023 \\
EPEX-DE & 01.01.2022 - 31.05.2023 \\
EPEX-BE & 04.01.2015 - 31.12.2016 \\
EPEX-FR & 04.01.2015 - 31.12.2016 \\
NP      & 27.12.2016 - 24.12.2018
\end{tabular}
\caption{Test period of every dataset considered}
\label{table:test_periods}
\end{table}

The LEAR model will be evaluated using four different calibration windows: two long and two short. According to \cite{lago2021forecasting}, the calibration windows for the EPEX-BE, EPEX-FR, and NP datasets will be 56, 84, 1092, and 1456 days, corresponding to 8 weeks, 12 weeks, 3 years, and 4 years, respectively. Given the higher volatility and more pronounced behavioral changes in the OMIE-SP and EPEX-DE datasets, shorter long windows of 56, 84, 364, and 728 days will be used. Furthermore, scenarios where the entire dataset is used for training will also be considered. This approach is particularly relevant for adaptive standardization, as the series seems to be comparable across all instances, which implies that the use of fixed calibration windows may be unnecessary if all changes are effectively captured by the parameters $\mu_d$ and $\sigma_d$. This study extends the work of \cite{lago2021forecasting} by including this comparison to critically assess the necessity for calibration windows in traditional normalisation schemes. For the DNN model, four different sets of hyperparameter optimizations are conducted, with the comparative results of these configurations examined. If adaptive standardization is not implemented, a calibration window of 4 years is employed for the EPEX-BE, EPEX-FR, and NP datasets, and 3 years for the OMIE-SP and EPEX-DE datasets. When adaptive standardization is applied, all available training data is used.\\

For the outlier filtering process, we consider $\kappa = 10$ as our intention is to filter out only prices that were really atypical and disproportionate to those of the last few days. For example, in Figure \ref{fig:market_prices} it can be seen near July 2023 in EPEX-BE and EPEX-FR clearly negative and atypical prices. In these same series, at the end of 2016, prices above 500 \euro/MWh can be seen which are contrary to the dynamics observed in the rest of the series. These are clear examples of outliers that we want to mitigate. Probably better results could be obtained with other values of $\theta$, as it actually is an hyperparameter in the methodology.\\

\subsection{Evaluation metrics}\label{sec:evaluation_metrics}

Four accepted metrics in the forecasting literature are used to measure the quality of outcomes:

\begin{itemize}
\item MAE $= \dfrac{1}{24N_d} \displaystyle \sum_{d=1}^{N_d} \sum_{h=1}^{24} \vert p_d^h - \hat{p}_d^h \vert$
\item RMSE $= \sqrt{\dfrac{1}{24N_d} \displaystyle \sum_{d=1}^{N_d} \sum_{h=1}^{24} \left( p_d^h - \hat{p}_d^h \right)^2}$
\item sMAPE $= \dfrac{1}{24N_d}\displaystyle \sum_{d=1}^{N_d} \sum_{h=1}^{24} 2 \dfrac{\vert p_d^h - \hat{p}_d^h \vert}{\vert p_d^h \vert + \vert\hat{p}_d^h \vert}$
\item rMAE $= \dfrac{\dfrac{1}{24N_d} \displaystyle \sum_{d=1}^{N_d} \sum_{h=1}^{24} \vert p_d^h - \hat{p}_d^h \vert}{\dfrac{1}{24N_d} \displaystyle \sum_{d=1}^{N_d} \sum_{h=1}^{24} \vert p_d^h - p_d^{h, \, \text{naive}} \vert}$
\end{itemize}
where $N_d$ is the number of days in the test set and $p_d^h$, $\hat{p}_d^h$, $p_d^{h, \, \text{naive}}$ are the current price, the prediction and the prediction by a naive model of the hour $h$ on day $d$, respectively. In the EPF field, is common to consider $p_d^{h, \, \text{naive}} = p_{d-7}^h$, so weekly effects are captured.\\

\subsection{Forecasting procedure}
For each dataset and for each day D+1 of the test period to be forecasted the process is as follows:

\begin{enumerate}
    \item Take as training data set the one given by the corresponding calibration window. If no calibration window is considered, take all data prior to the day to be predicted as training data.
    \item If adaptive standardisation is applied, transform the price series by filtering outliers as explained in Section \ref{sec:new_model}.
    \item Transform the price series and the explanatory variables using the selected normalisation process. Either adaptive standardisation or median-arcsinh.
    \item Select and train the learning algorithm: LEAR or DNN.
    \item Predict all hours of the day D+1.
    \item Undo the selected normalisation scheme.
\end{enumerate}

When all days of the test period have been forecasted, the metrics presented in Section \ref{sec:evaluation_metrics} are computed.\\

In addition, results have been computed without filtering outliers in the case of adaptive standardisation to show the importance of this step (\ref{ap:results_no_filter}). Also, to show that this step alone is not the one that produces considerable improvements, the results of the filtered series have been computed by applying the median-arcsinh transformation instead of applying the adaptive standardisation (\ref{ap:results_filter}).\\

\subsection{Results}

Tables \ref{table:metrics_LEAR} and \ref{table:metrics_DNN} show the evaluation metrics for the mentioned datasets for the algorithms LEAR and DNN, respectively. We will adopt the notation ``AS'' before each model name to indicate that it incorporates the adaptive standardisation approach.\\

\begin{table}[h]
\centering
\resizebox{\textwidth}{!}{
\begin{tabular}{c|c|lllll|lllll}
\multicolumn{1}{l|}{} & \multicolumn{1}{l|}{} & \multicolumn{5}{c|}{\textbf{LEAR}} & \multicolumn{5}{c}{\textbf{ASLEAR}} \\ \hline
\textbf{Market} & \textbf{Metrics} & \multicolumn{1}{c}{\textbf{56}} & \multicolumn{1}{c}{\textbf{84}} & \multicolumn{1}{c}{\textbf{\begin{tabular}[c]{@{}c@{}}364/\\ 1092\end{tabular}}} & \multicolumn{1}{c}{\textbf{\begin{tabular}[c]{@{}c@{}}728/\\ 1456\end{tabular}}} & \multicolumn{1}{c|}{\textbf{All}} & \multicolumn{1}{c}{\textbf{56}} & \multicolumn{1}{c}{\textbf{84}} & \multicolumn{1}{c}{\textbf{\begin{tabular}[c]{@{}c@{}}364/\\ 1092\end{tabular}}} & \multicolumn{1}{c}{\textbf{\begin{tabular}[c]{@{}c@{}}728/\\ 1456\end{tabular}}} & \multicolumn{1}{c}{\textbf{All}} \\ \hline
\multirow{4}{*}{\textbf{\begin{tabular}[c]{@{}c@{}}OMIE\\ SP\end{tabular}}} & \textbf{MAE} & 21,03 & 20,86 & 19,40 & 19,46 & 20,69 & 21,75 & 20,35 & 18,68 & 18,48 & \textbf{18,27} \\
 & \textbf{RMSE} & 30,66 & 31,09 & 27,96 & 27,57 & 29,28 & 32,31 & 29,26 & 26,78 & 26,18 & \textbf{25,93} \\
 & \textbf{sMAPE} & 0,23 & 0,23 & 0,22 & 0,22 & 0,22 & 0,23 & 0,22 & \textbf{0,21} & \textbf{0,21} & \textbf{0,21} \\
 & \textbf{rMAE} & 0,55 & 0,54 & 0,51 & 0,51 & 0,54 & 0,57 & 0,53 & 0,49 & \textbf{0,48} & \textbf{0,48} \\ \hline
\multirow{4}{*}{\textbf{\begin{tabular}[c]{@{}c@{}}EPEX\\ DE\end{tabular}}} & \textbf{MAE} & 31,42 & 30,45 & 30,67 & 28,54 & 31,13 & 29,72 & 28,76 & 26,47 & 25,99 & \textbf{25,65} \\
 & \textbf{RMSE} & 47,31 & 44,99 & 42,52 & 40,60 & 44,60 & 44,67 & 42,69 & 39,14 & 38,65 & \textbf{38,11} \\
 & \textbf{sMAPE} & 0,25 & 0,25 & 0,25 & 0,23 & 0,24 & 0,24 & 0,23 & 0,22 & 0,22 & \textbf{0,21} \\
 & \textbf{rMAE} & 0,44 & 0,42 & 0,43 & 0,40 & 0,43 & 0,41 & 0,40 & 0,37 & \textbf{0,36} & \textbf{0,36} \\ \hline
\multirow{4}{*}{\textbf{\begin{tabular}[c]{@{}c@{}}EPEX\\ BE\end{tabular}}} & \textbf{MAE} & 7,03 & 6,92 & \textbf{6,45} & \textbf{6,45} & 6,54 & 8,30 & 7,83 & 6,85 & 6,84 & 6,84 \\
 & \textbf{RMSE} & \textbf{16,16} & 16,19 & 16,51 & 16,42 & 16,36 & 25,92 & 26,16 & 17,16 & 17,24 & 17,24 \\
 & \textbf{sMAPE} & 0,17 & 0,16 & 0,16 & 0,16 & 0,16 & 0,17 & 0,17 & 0,16 & 0,16 & \textbf{0,15} \\
 & \textbf{rMAE} & 0,69 & 0,68 & \textbf{0,64} & \textbf{0,64} & \textbf{0,64} & 0,82 & 0,77 & 0,67 & 0,67 & 0,67 \\ \hline
\multirow{4}{*}{\textbf{\begin{tabular}[c]{@{}c@{}}EPEX\\ FR\end{tabular}}} & \textbf{MAE} & 4,82 & 4,62 & 4,21 & 4,28 & 4,37 & 4,86 & 4,69 & 4,15 & \textbf{4,14} & 4,19 \\
 & \textbf{RMSE} & \textbf{10,83} & 11,49 & 11,67 & 11,67 & 11,67 & 14,30 & 14,56 & 12,17 & 12,24 & 12,64 \\
 & \textbf{sMAPE} & 0,14 & 0,13 & 0,13 & 0,13 & 0,14 & 0,13 & 0,13 & \textbf{0,12} & \textbf{0,12} & \textbf{0,12} \\
 & \textbf{rMAE} & 0,66 & 0,63 & \textbf{0,57} & 0,58 & 0,60 & 0,66 & 0,64 & \textbf{0,57} & \textbf{0,57} & \textbf{0,57} \\ \hline
\multirow{4}{*}{\textbf{NP}} & \textbf{MAE} & 2,02 & 1,96 & 1,96 & 1,96 & \textbf{1,92} & 2,61 & 2,47 & 2,01 & 1,98 & 1,97 \\
 & \textbf{RMSE} & 3,76 & 3,73 & 3,56 & 3,57 & \textbf{3,53} & 4,88 & 4,79 & 3,83 & 3,80 & 3,80 \\
 & \textbf{sMAPE} & \textbf{0,06} & \textbf{0,06} & \textbf{0,06} & \textbf{0,06} & \textbf{0,06} & 0,08 & 0,07 & \textbf{0,06} & \textbf{0,06} & \textbf{0,06} \\
 & \textbf{rMAE} & 0,49 & 0,47 & 0,47 & 0,47 & \textbf{0,46} & 0,63 & 0,60 & 0,49 & 0,48 & 0,48
\end{tabular}
}
\caption{Evaluation metrics the LEAR and ASLEAR models for every dataset}
\label{table:metrics_LEAR}
\end{table}

\begin{table}[h]
\centering
\resizebox{\textwidth}{!}{
\begin{tabular}{c|c|llll|llll}
\multicolumn{1}{l|}{} & \multicolumn{1}{l|}{} & \multicolumn{4}{c|}{\textbf{DNN}} & \multicolumn{4}{c}{\textbf{ASDNN}} \\ \hline
\textbf{Market} & \textbf{Metrics} & \multicolumn{1}{c}{\textbf{1}} & \multicolumn{1}{c}{\textbf{2}} & \multicolumn{1}{c}{\textbf{3}} & \multicolumn{1}{c|}{\textbf{4}} & \multicolumn{1}{c}{\textbf{1}} & \multicolumn{1}{c}{\textbf{2}} & \multicolumn{1}{c}{\textbf{3}} & \multicolumn{1}{c}{\textbf{4}} \\ \hline
\multirow{4}{*}{\textbf{\begin{tabular}[c]{@{}c@{}}OMIE\\ SP\end{tabular}}} & \textbf{MAE} & 22,38 & 19,24 & 23,37 & 20,79 & 18,19 & 18,10 & 17,87 & \textbf{17,67} \\
 & \textbf{RMSE} & 31,82 & 27,46 & 32,96 & 30,15 & 26,01 & 26,12 & 25,31 & \textbf{25,11} \\
 & \textbf{sMAPE} & 0,23 & \textbf{0,21} & 0,24 & 0,22 & \textbf{0,21} & \textbf{0,21} & \textbf{0,21} & \textbf{0,21} \\
 & \textbf{rMAE} & 0,58 & 0,50 & 0,61 & 0,54 & 0,47 & 0,47 & 0,47 & \textbf{0,46} \\ \hline
\multirow{4}{*}{\textbf{\begin{tabular}[c]{@{}c@{}}EPEX\\ DE\end{tabular}}} & \textbf{MAE} & 27,21 & 25,61 & 26,18 & 26,83 & 26,28 & 25,92 & \textbf{24,35} & 26,26 \\
 & \textbf{RMSE} & 39,17 & 37,32 & 38,14 & 38,69 & 38,75 & 38,37 & \textbf{36,11} & 38,70 \\
 & \textbf{sMAPE} & 0,22 & \textbf{0,21} & 0,22 & 0,22 & \textbf{0,21} & \textbf{0,21} & \textbf{0,21} & \textbf{0,21} \\
 & \textbf{rMAE} & 0,38 & 0,36 & 0,36 & 0,37 & 0,36 & 0,36 & \textbf{0,34} & 0,36 \\ \hline
\multirow{4}{*}{\textbf{\begin{tabular}[c]{@{}c@{}}EPEX\\ BE\end{tabular}}} & \textbf{MAE} & 6,29 & \textbf{6,22} & 6,46 & 6,42 & 6,39 & 6,58 & 6,65 & 6,44 \\
 & \textbf{RMSE} & 16,13 & 16,24 & 16,63 & 16,55 & \textbf{15,89} & 16,48 & 16,55 & 16,21 \\
 & \textbf{sMAPE} & \textbf{0,15} & \textbf{0,15} & \textbf{0,15} & \textbf{0,15} & \textbf{0,15} & \textbf{0,15} & 0,16 & \textbf{0,15} \\
 & \textbf{rMAE} & 0,62 & \textbf{0,61} & 0,64 & 0,63 & 0,63 & 0,65 & 0,65 & 0,63 \\ \hline
\multirow{4}{*}{\textbf{\begin{tabular}[c]{@{}c@{}}EPEX\\ FR\end{tabular}}} & \textbf{MAE} & 4,25 & 4,24 & 4,23 & 4,12 & \textbf{4,08} & 4,11 & 4,16 & 4,17 \\
 & \textbf{RMSE} & 11,94 & 12,14 & 11,89 & 12,01 & \textbf{11,16} & 12,04 & 11,90 & 11,71 \\
 & \textbf{sMAPE} & \textbf{0,12} & \textbf{0,12} & \textbf{0,12} & \textbf{0,12} & \textbf{0,12} & \textbf{0,12} & \textbf{0,12} & \textbf{0,12} \\
 & \textbf{rMAE} & 0,58 & 0,58 & 0,58 & \textbf{0,56} & \textbf{0,56} & \textbf{0,56} & 0,57 & 0,57 \\ \hline
\multirow{4}{*}{\textbf{NP}} & \textbf{MAE} & 2,11 & 1,83 & 1,96 & 1,84 & 1,74 & 1,76 & \textbf{1,73} & 1,77 \\
 & \textbf{RMSE} & 3,95 & 3,47 & 3,67 & 3,50 & \textbf{3,37} & 3,48 & 3,46 & 3,58 \\
 & \textbf{sMAPE} & 0,06 & \textbf{0,05} & 0,06 & \textbf{0,05} & \textbf{0,05} & \textbf{0,05} & \textbf{0,05} & \textbf{0,05} \\
 & \textbf{rMAE} & 0,51 & 0,44 & 0,48 & 0,45 & \textbf{0,42} & \textbf{0,42} & \textbf{0,42} & 0,43
\end{tabular}
}
\caption{Evaluation metrics the DNN and ASDNN models for every dataset\protect\footnotemark}
\label{table:metrics_DNN}
\end{table}

Looking at the performance of the LEAR model it can be seen that larger windows produce better results. However, there is a point where this larger calibration window stops producing improvements or even worsens the results, except for the NP market. This makes sense, since in view of this model not all observations are comparable to each other. The ASLEAR model does have this feature, always taking advantage of all the available information. It can be seen how the best results for this case are always considering all the data, except for the EPEX-FR market, where they are practically the same. However, the results with small calibration windows are not positive, being very distant from the rest of the configurations. However, this is inline with the intended used of the methodology. Comparing LEAR and ASLEAR, the biggest differences appear in the OMIE-SP and EPEX-DE markets, the markets that deal with the least stable periods. For the former, comparing the best models, the predictions are 6\% better in MAE, and for the latter dataset 10\% better, very remarkable figures. For the rest of the datasets the results are very similar comparing the best models in each case. These datasets show little change in behaviour, so the adaptive standardisation is actually producing very similar results to what a static standardisation would be and, in fact, what is observed are the differences in performance of two different static standardisations. In any case, the results are slightly better for LEAR in the case of EPEX-BE and NP and for ASLEAR in the case of EPEX-FR.\\

\footnotetext{It is important to note that DNN and ASDNN are not comparable for a hyperparameter configuration labelled with the same number. It simply means four different configurations for each case.}

With regard to the DNN and ASDNN models, the first point to be made concerns the volatility of the results. Neural networks, due to their hyperparameter configuration and the initialisation of the network weights, can give very different results for the same network architecture. In the case of the DNN this variation is very clear, producing highly diverse results for each dataset. The best examples are seen in the OMIE-SP, EPEX-DE and NP markets, which are also the most volatile markets, at least compared to the other two. For the ASDNN model this variation in results is minimal, seeing very stable results across all configurations for all datasets, perhaps with the exception of EPEX-DE. This property is important, as relatively good performance is ensured without the need for multiple runs or combining different networks. Regarding the performance on each dataset, the trend of the LEAR and ASLEAR models is somewhat observed. For OMIE-SP the difference is even more noticeable, where all ASDNN networks are better than the best DNN network. A significant difference is also seen in the case of EPEX-DE, as expected. The DNN model is still better than the ASDNN for EPEX-BE, although the results are still very close. For EPEX-FR the ASDNN model is better for all networks, except for one DNN network which performs at the same level. Finally, in the case of the NP market, we observe a separation that was not present between the LEAR and ASLEAR, perhaps because of the now existing ability to capture non-linear behaviour. Under these conditions, the ASDNN model performs better for all networks. \\

An easy improvement, as shown in \cite{lago2021forecasting}, can be achieved by creating ensembles of the different predictions. Normally, the simplest schemes in this sense already obtain very good results that are hard to beat. Thus, for each group of models (LEAR, ASLEAR, DNN and ASDNN) the mean of all the predictions of each group is considered as a new predictor. In the case of LEAR, the case in which all the observations available for training are considered is excluded, since it is not included in the state-of-the-art model of \cite{lago2021forecasting} either. In addition, for ASLEAR two variants, 1 (2), are considered, where the 54 and 86 day calibration windows are not (are) considered because of their poor individual performance. Table \ref{table:metrics_ensembles} presents the results, comparing each ensemble with the best individual model of each group.\\

\begin{table}[h]
\centering
\resizebox{\textwidth}{!}{
\begin{tabular}{c|c|cc|ccc|cc|cc}
\textbf{Market} & \textbf{Metrics} & \textbf{\begin{tabular}[c]{@{}c@{}}Ens.\\ LEAR\end{tabular}} & \textbf{\begin{tabular}[c]{@{}c@{}}Best\\ LEAR\end{tabular}} & \textbf{\begin{tabular}[c]{@{}c@{}}Ens.$_1$\\ ASLEAR\end{tabular}} & \textbf{\begin{tabular}[c]{@{}c@{}}Ens.$_2$\\ ASLEAR\end{tabular}} & \textbf{\begin{tabular}[c]{@{}c@{}}Best\\ ASLEAR\end{tabular}} & \textbf{\begin{tabular}[c]{@{}c@{}}Ens.\\ DNN\end{tabular}} & \textbf{\begin{tabular}[c]{@{}c@{}}Best\\ DNN\end{tabular}} & \textbf{\begin{tabular}[c]{@{}c@{}}Ens.\\ ASDNN\end{tabular}} & \textbf{\begin{tabular}[c]{@{}c@{}}Best\\ ASDNN\end{tabular}} \\ \hline
\multirow{4}{*}{\textbf{\begin{tabular}[c]{@{}c@{}}OMIE\\ SP\end{tabular}}} & \textbf{MAE} & 18,19 & 19,40 & 18,11 & 18,16 & 18,27 & 19,60 & 19,24 & \textbf{17,29} & 17,67 \\
 & \textbf{RMSE} & 26,16 & 27,96 & 25,86 & 26,20 & 25,93 & 28,60 & 27,46 & \textbf{24,70} & 25,11 \\
 & \textbf{sMAPE} & 0,20 & 0,22 & 0,21 & \textbf{0,20} & 0,21 & 0,21 & 0,21 & \textbf{0,20} & 0,21 \\
 & \textbf{rMAE} & 0,47 & 0,51 & 0,47 & 0,47 & 0,48 & 0,51 & 0,50 & \textbf{0,45} & 0,46 \\ \hline
\multirow{4}{*}{\textbf{\begin{tabular}[c]{@{}c@{}}EPEX\\ DE\end{tabular}}} & \textbf{MAE} & 26,24 & 28,54 & 25,52 & 25,41 & 25,65 & \textbf{23,69} & 25,61 & 23,91 & 24,35 \\
 & \textbf{RMSE} & 37,78 & 40,60 & 37,99 & 37,97 & 38,11 & \textbf{34,61} & 37,32 & 35,22 & 36,11 \\
 & \textbf{sMAPE} & 0,22 & 0,23 & 0,21 & 0,21 & 0,21 & \textbf{0,20} & 0,21 & \textbf{0,20} & 0,21 \\
 & \textbf{rMAE} & 0,36 & 0,40 & 0,35 & 0,35 & 0,36 & \textbf{0,33} & 0,36 & \textbf{0,33} & 0,34 \\ \hline
\multirow{4}{*}{\textbf{\begin{tabular}[c]{@{}c@{}}EPEX\\ BE\end{tabular}}} & \textbf{MAE} & 6,22 & 6,45 & 6,82 & 6,96 & 6,84 & \textbf{6,06} & 6,22 & 6,31 & 6,39 \\
 & \textbf{RMSE} & 15,85 & 16,42 & 17,15 & 18,50 & 17,24 & 16,14 & 16,24 & 16,07 & \textbf{15,89} \\
 & \textbf{sMAPE} & 0,15 & 0,16 & 0,15 & 0,15 & 0,15 & \textbf{0,14} & 0,15 & 0,15 & 0,15 \\
 & \textbf{rMAE} & 0,61 & 0,64 & 0,67 & 0,69 & 0,67 & \textbf{0,60} & 0,61 & 0,62 & 0,63 \\ \hline
\multirow{4}{*}{\textbf{\begin{tabular}[c]{@{}c@{}}EPEX\\ FR\end{tabular}}} & \textbf{MAE} & 4,04 & 4,21 & 4,14 & 4,17 & 4,14 & 4,00 & 4,12 & \textbf{3,94} & 4,08 \\
 & \textbf{RMSE} & 10,86 & 11,67 & 12,34 & 12,63 & 12,24 & 11,87 & 12,01 & 11,53 & \textbf{11,16} \\
 & \textbf{sMAPE} & 0,12 & 0,13 & \textbf{0,11} & \textbf{0,11} & 0,12 & \textbf{0,11} & 0,12 & \textbf{0,11} & 0,12 \\
 & \textbf{rMAE} & 0,55 & 0,57 & 0,57 & 0,57 & 0,57 & 0,55 & 0,56 & \textbf{0,54} & 0,56 \\ \hline
\multirow{4}{*}{\textbf{NP}} & \textbf{MAE} & 1,75 & 1,96 & 1,98 & 2,02 & 1,97 & 1,74 & 1,83 & \textbf{1,64} & 1,73 \\
 & \textbf{RMSE} & 3,39 & 3,56 & 3,80 & 3,92 & 3,80 & 3,44 & 3,47 & \textbf{3,33} & 3,46 \\
 & \textbf{sMAPE} & \textbf{0,05} & 0,06 & 0,06 & 0,06 & 0,06 & \textbf{0,05} & \textbf{0,05} & \textbf{0,05} & \textbf{0,05} \\
 & \textbf{rMAE} & 0,42 & 0,47 & 0,48 & 0,49 & 0,48 & 0,42 & 0,44 & \textbf{0,40} & 0,42
\end{tabular}
}
\caption{Evaluation metrics for the ensembles of the different models computed}
\label{table:metrics_ensembles}
\end{table}

When ensembles are computed, methods that make use of adaptive standardisation do not improve as much as those that do not. The variance of the individual models when adaptive standardisation is not applied is by itself a negative feature when a single model is to be used. Nevertheless, when several models are mixed together they take into account very diverse behaviour, leading to large improvements in the metrics considered. In particular, in the case of OMIE-SP, the ASLEAR and ASDNN ensembles are better than their respective non-adaptive versions. However, the improvement over the best single model is much larger for the non-adaptive schemes. For EPEX-DE the ASLEAR ensembles remain better, but the DNN ensemble slightly outperforms the ASDNN ensemble in MAE, although they maintain the same level of rMAE. For the EPEX-BE market there was already a better performance of the non-adaptive models and this trend is also maintained in this case. In the French case, EPEX-FR, the ASLEAR ensembles do not produce any improvement over the best individual model, while for LEAR they do. Still, the ASDNN ensemble is the best model over the DNN combination. Finally, the same behaviour is observed for NP as for France, where ASDNN is again the best model. Ultimately, the ASDNN ensembles are generally the best models, but the improvement over the best model is not as great as for the non-adaptive versions. While this may seem like a negative feature for adaptive standardisation, it actually indicates that very similar performance to the combination of non-adaptive models can be achieved using a single model to which adaptive standardisation has been applied.\\

\begin{figure}[htp]
    \centering
    \includegraphics[scale=0.45]{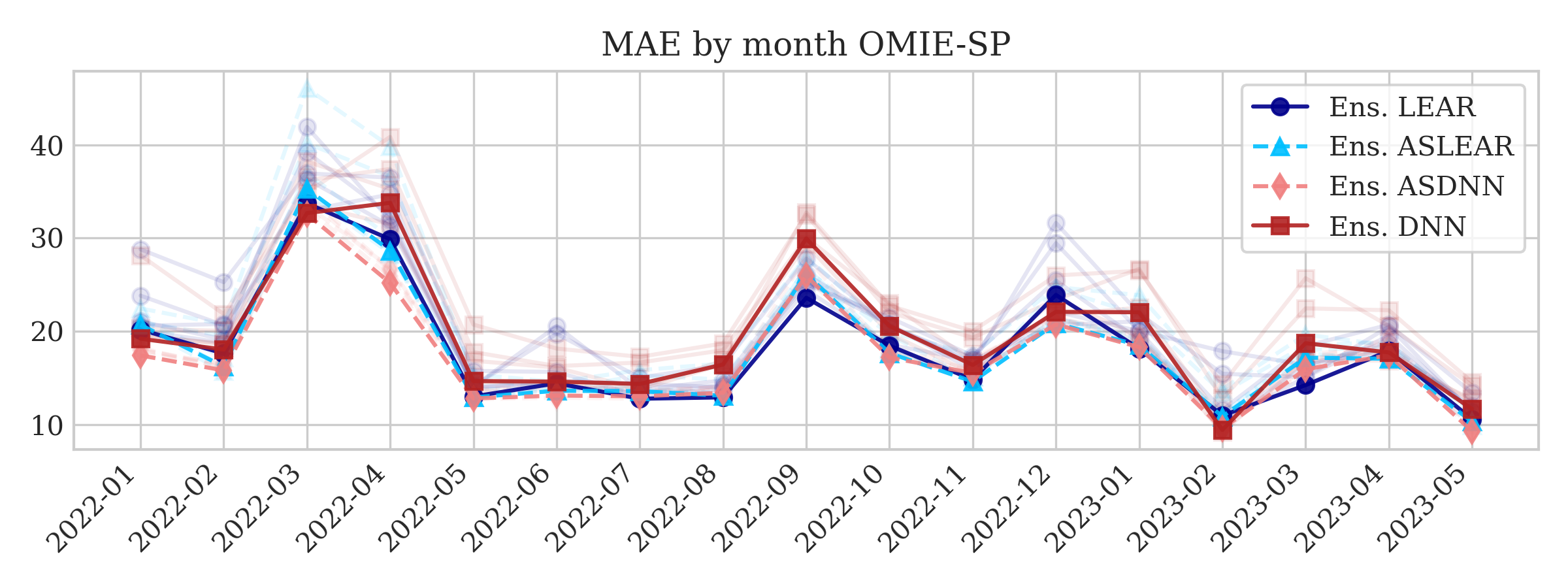}
    \includegraphics[scale=0.45]{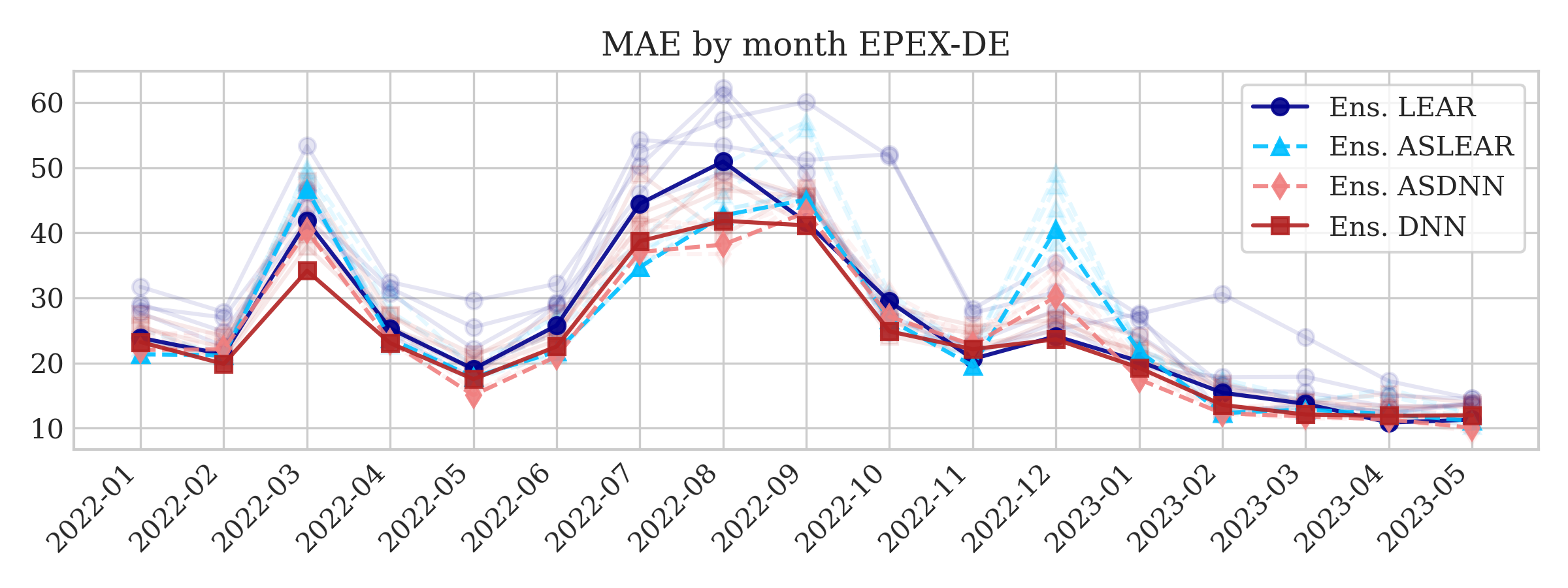}
    \includegraphics[scale=0.45]{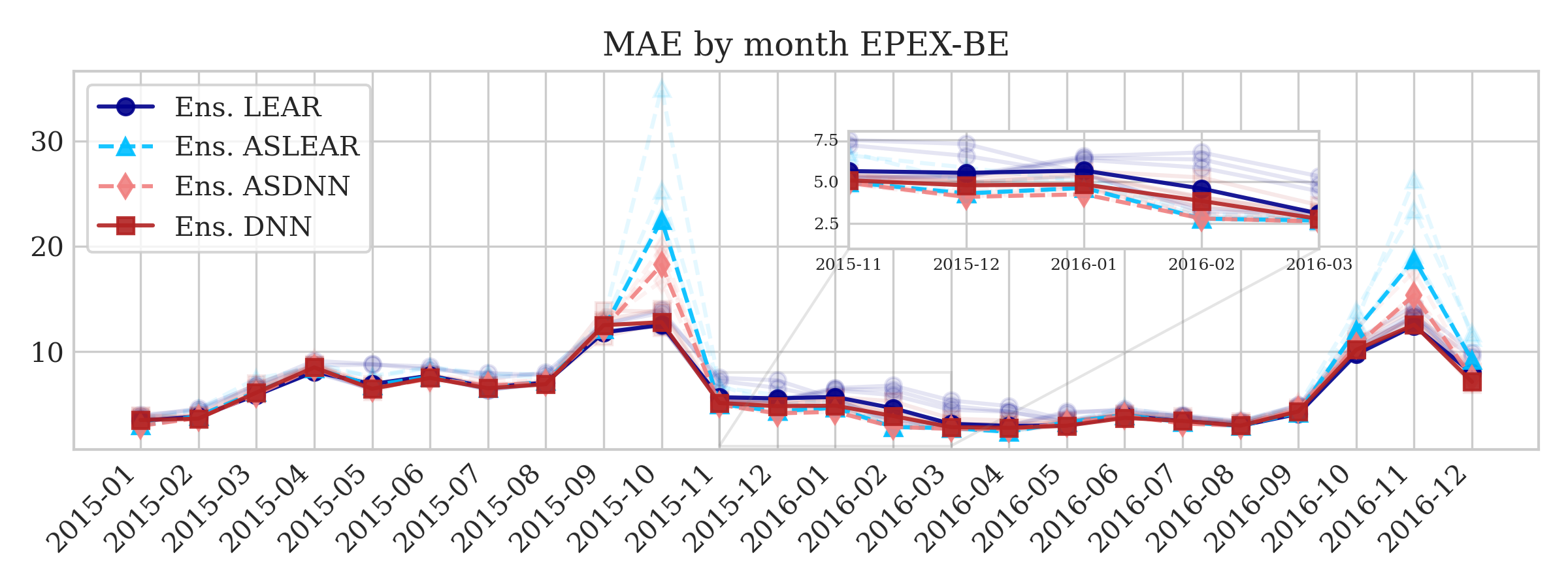}
    \includegraphics[scale=0.45]{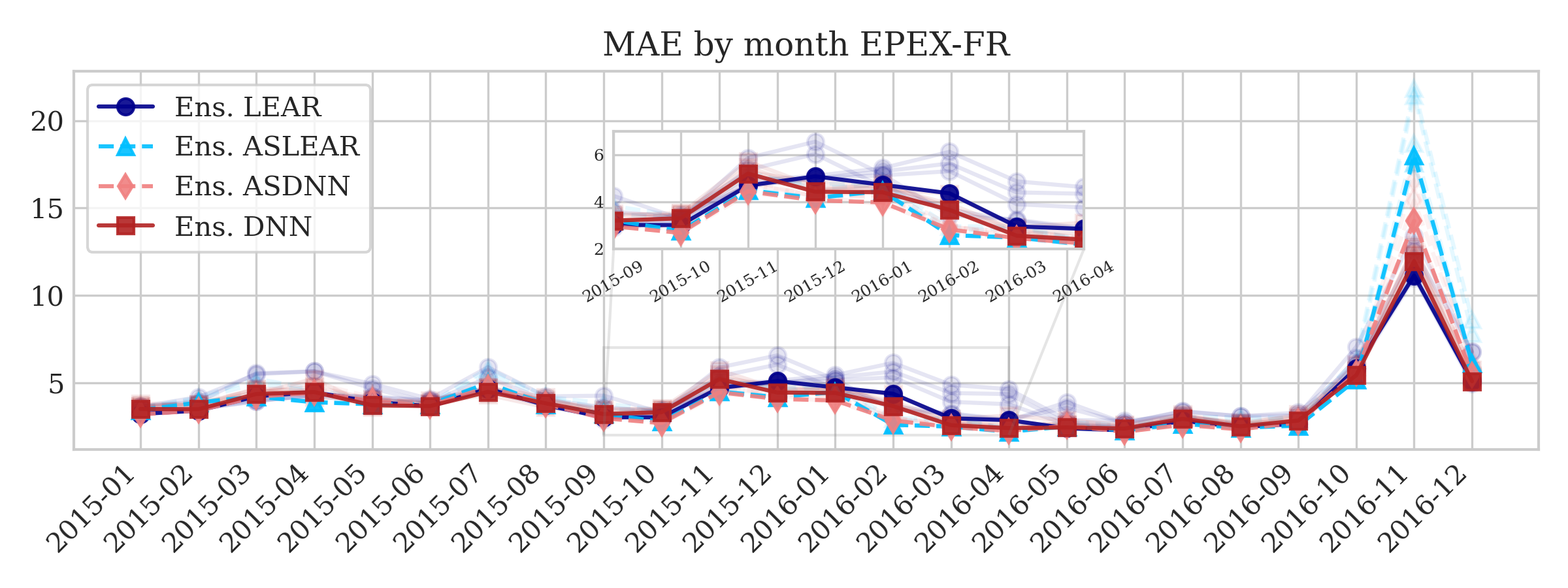}
    \includegraphics[scale=0.45]{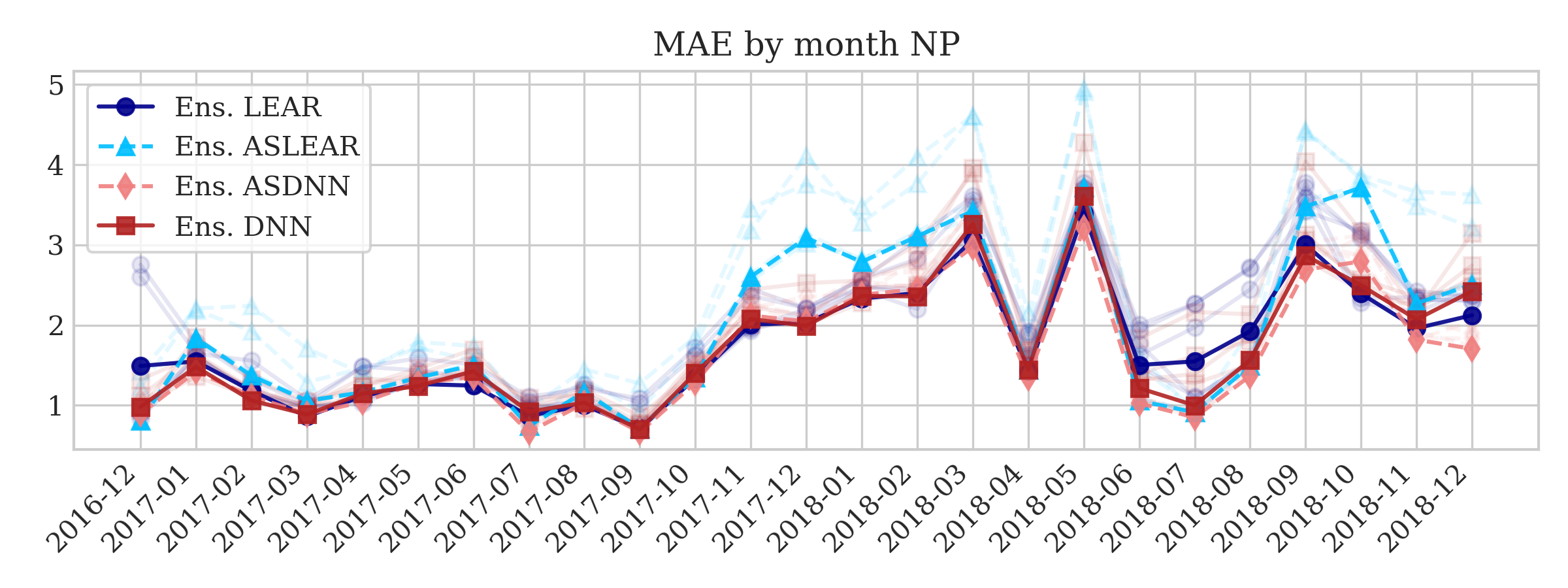}
    \caption{MAE per month and per dataset for each model set. Each individual model is shown in translucent form.}
    \label{fig:maes_by_month}
\end{figure}

We believe that a good practice for evaluating models, which is commonly ignored, is the dynamic evaluation of some metric. For instance, the MAE per month of each of the models. Figure \ref{fig:maes_by_month} compares the ensembles with each other for each of the datasets on a month-by-month basis. In the following, the ensemble of the ASLEAR model set never considers the 56 and 84 day windows.\\

We can now appreciate where the advantages of each method come from: whether they are period-specific or consistent over time. In the case of OMIE-SP it can be seen that the adaptive methods perform better over the entire period evaluated. This could be suspected as the gain in this market is very significant compared to the other markets. In the EPEX-DE market there are mixed situations where the adaptive and non-adaptive versions exchange the best model position. There is also a significant jump in quality between using the (AS)LEAR model and the (AS)DNN model. The Belgian and French markets are very similar. The main differences are given by those MAE peaks, which are coincidentally periods where very atypical prices stand out. Probably with a better treatment of outliers (e.g. by not setting such a high $\theta$) these problems would be alleviated. Other than that, performance is roughly equivalent, except for a period in late 2015 and early 2016 where in both markets the adaptive versions are slightly better. For NP, the performance of the ASLEAR ensemble was by far the worst of all (Table \ref{table:metrics_ensembles}), as it was the only model that did not improve when averaging over the individual models. It can now be seen that this is mainly determined by a period from November 2017 to March 2018.\\

Another property that can be observed from these plots is that the adaptive and non-adaptive versions capture different behaviours. Just as it made sense to combine models from different calibration windows or different hyperparameter settings, it makes sense to combine the adaptive and non-adaptive versions. These combinations are denoted by LEAR-ASLEAR and DNN-ASDNN and are constructed by making the arithmetic mean of the two ensembles involved. Table \ref{table:metrics_ensembles_2} shows the results.\\

\begin{table}[h]
\centering
\resizebox{\textwidth}{!}{
\begin{tabular}{c|c|ccc|ccc}
\textbf{Market} & \textbf{Metrics} & \textbf{\begin{tabular}[c]{@{}c@{}}Ens.\\ LEAR\end{tabular}} & \textbf{\begin{tabular}[c]{@{}c@{}}Ens.\\ ASLEAR\end{tabular}} & \textbf{\begin{tabular}[c]{@{}c@{}}LEAR\\ ASLEAR\end{tabular}} & \textbf{\begin{tabular}[c]{@{}c@{}}Ens.\\ DNN\end{tabular}} & \textbf{\begin{tabular}[c]{@{}c@{}}Ens.\\ ASDNN\end{tabular}} & \textbf{\begin{tabular}[c]{@{}c@{}}DNN\\ ASDNN\end{tabular}} \\ \hline
\multirow{4}{*}{\textbf{\begin{tabular}[c]{@{}c@{}}OMIE\\ SP\end{tabular}}} & \textbf{MAE} & 18,19 & 18,11 & 17,30 & 19,60 & \textbf{17,29} & 17,33 \\
 & \textbf{RMSE} & 26,16 & 25,86 & 24,76 & 28,60 & \textbf{24,70} & 25,23 \\
 & \textbf{sMAPE} & 0,20 & 0,21 & \textbf{0,20} & 0,21 & \textbf{0,20} & \textbf{0,20} \\
 & \textbf{rMAE} & 0,47 & 0,47 & \textbf{0,45} & 0,51 & \textbf{0,45} & \textbf{0,45} \\ \hline
\multirow{4}{*}{\textbf{\begin{tabular}[c]{@{}c@{}}EPEX\\ DE\end{tabular}}} & \textbf{MAE} & 26,24 & 25,52 & 23,84 & 23,69 & 23,91 & \textbf{22,10} \\
 & \textbf{RMSE} & 37,78 & 37,99 & 34,89 & 34,61 & 35,22 & \textbf{32,47} \\
 & \textbf{sMAPE} & 0,22 & 0,21 & 0,20 & 0,20 & 0,20 & \textbf{0,19} \\
 & \textbf{rMAE} & 0,36 & 0,35 & 0,33 & 0,33 & 0,33 & \textbf{0,31} \\ \hline
\multirow{4}{*}{\textbf{\begin{tabular}[c]{@{}c@{}}EPEX\\ BE\end{tabular}}} & \textbf{MAE} & 6,22 & 6,82 & 6,23 & 6,06 & 6,31 & \textbf{5,95} \\
 & \textbf{RMSE} & 15,85 & 17,15 & 15,85 & 16,14 & 16,07 & \textbf{15,84} \\
 & \textbf{sMAPE} & 0,15 & 0,15 & \textbf{0,14} & \textbf{0,14} & 0,15 & \textbf{0,14} \\
 & \textbf{rMAE} & 0,61 & 0,67 & 0,61 & 0,60 & 0,62 & \textbf{0,59} \\ \hline
\multirow{4}{*}{\textbf{\begin{tabular}[c]{@{}c@{}}EPEX\\ FR\end{tabular}}} & \textbf{MAE} & 4,04 & 4,14 & 3,90 & 4,00 & 3,94 & \textbf{3,81} \\
 & \textbf{RMSE} & 10,86 & 12,34 & \textbf{11,38} & 11,87 & 11,53 & 11,54 \\
 & \textbf{sMAPE} & 0,12 & 0,11 & 0,11 & 0,11 & 0,11 & 0,11 \\
 & \textbf{rMAE} & 0,55 & 0,57 & 0,53 & 0,55 & 0,54 & \textbf{0,52} \\ \hline
\multirow{4}{*}{\textbf{NP}} & \textbf{MAE} & 1,75 & 1,98 & 1,74 & 1,74 & 1,64 & \textbf{1,60} \\
 & \textbf{RMSE} & 3,39 & 3,80 & 3,40 & 3,44 & 3,33 & \textbf{3,27} \\
 & \textbf{sMAPE} & \textbf{0,05} & 0,06 & \textbf{0,05} & \textbf{0,05} & \textbf{0,05} & \textbf{0,05} \\
 & \textbf{rMAE} & 0,42 & 0,48 & 0,42 & 0,42 & 0,40 & \textbf{0,39}
\end{tabular}
}
\caption{Evaluation metrics for the ensembles and the combination of the adaptive methods with the non-adaptive ones}
\label{table:metrics_ensembles_2}
\end{table}

For the OMIE-SP market, the difference between adaptive and non-adaptive versions is so large that there is no improvement when combining neural networks. However, when combining the linear models the improvement is very significant, reaching the level of the ASDNN ensemble. For the rest of the markets, the best model is given by the DNN-ASDNNN combination and, generally, by a wide margin with respect to the second best proposal. Therefore, although the individual models of the adaptive versions perform well on their own, and although combinations of models from the same methodology provide excellent performance, the real gain comes from using the two methodologies together, as they take into account properties that are captured by both techniques.\\

\subsection{Statistical testing}

It is important to analyze whether the difference between predictions from different models is statistically significant. In the context of EPF, the Diebold-Mariano test \citep{diebold2002comparing} is used for this purpose.\\

The Diebold-Mariano test evaluates the hypothesis that the expected value of $\Delta_{d, h}^{A, B}$ is zero, where $\Delta_{d, h}^{A, B} = L(\varepsilon_{d, h}^A) - L(\varepsilon_{d, h}^B)$. Here, $\varepsilon_{d, h}^Z = p_d^h - \hat{p}_d^h$ represents the prediction error of model $Z$ for day $d$ at hour $h$, and $L$ denotes the loss function used in the analysis. For this study, we consider the absolute loss as the loss function.\\

The statistic $\text{DM} = \sqrt{N}\dfrac{\hat{\mu}}{\hat{\sigma}}$ is computed, where $\hat{\mu}$ and $\hat{\sigma}$ represent the mean and standard deviation of $\Delta_{d, h}^{A, B}$, respectively, and $N$ is the number of observations in the test period. The test statistic asymptotically follows a standard normal distribution. In this way, the computation of the p-value associated with the test $$\begin{cases} 
H_0: &\mathbb{E}\left( \Delta_{d, h}^{A, B} \right) \leq 0 \\
H_1 : &\mathbb{E}\left( \Delta_{d, h}^{A, B} \right) > 0
\end{cases}$$ can be obtained.\\

If the null hypothesis is rejected, it indicates that the forecasts of model $B$ are statistically more accurate than those of model $A$.\\

It should be noted that the Diebold-Mariano test assumes that the observations are covariance stationary. Given that predictions are made for all hourly periods of day D+1, it is likely that this condition may not hold. To address this issue, there are two possible approaches. The first approach involves transforming the hourly time series into 24 daily series. This allows for separate tests to be performed for each hour of the day, resulting in 24 individual tests. This hourly univariate perspective enables a more detailed analysis of the model performance for each specific hour. Alternatively, a multivariate perspective can be adopted by considering 24-dimensional vectors representing the predictions for all hours of the day. In this case, a single test is conducted based on the norms of these vectors. This multivariate perspective provides a more concise and direct analysis, as the comparison is summarized in a single test statistic. This last version will be applied through the \texttt{epftoolbox}.

\subsubsection{Results}

\begin{figure}[h]
\centering
\includegraphics[width=\textwidth]{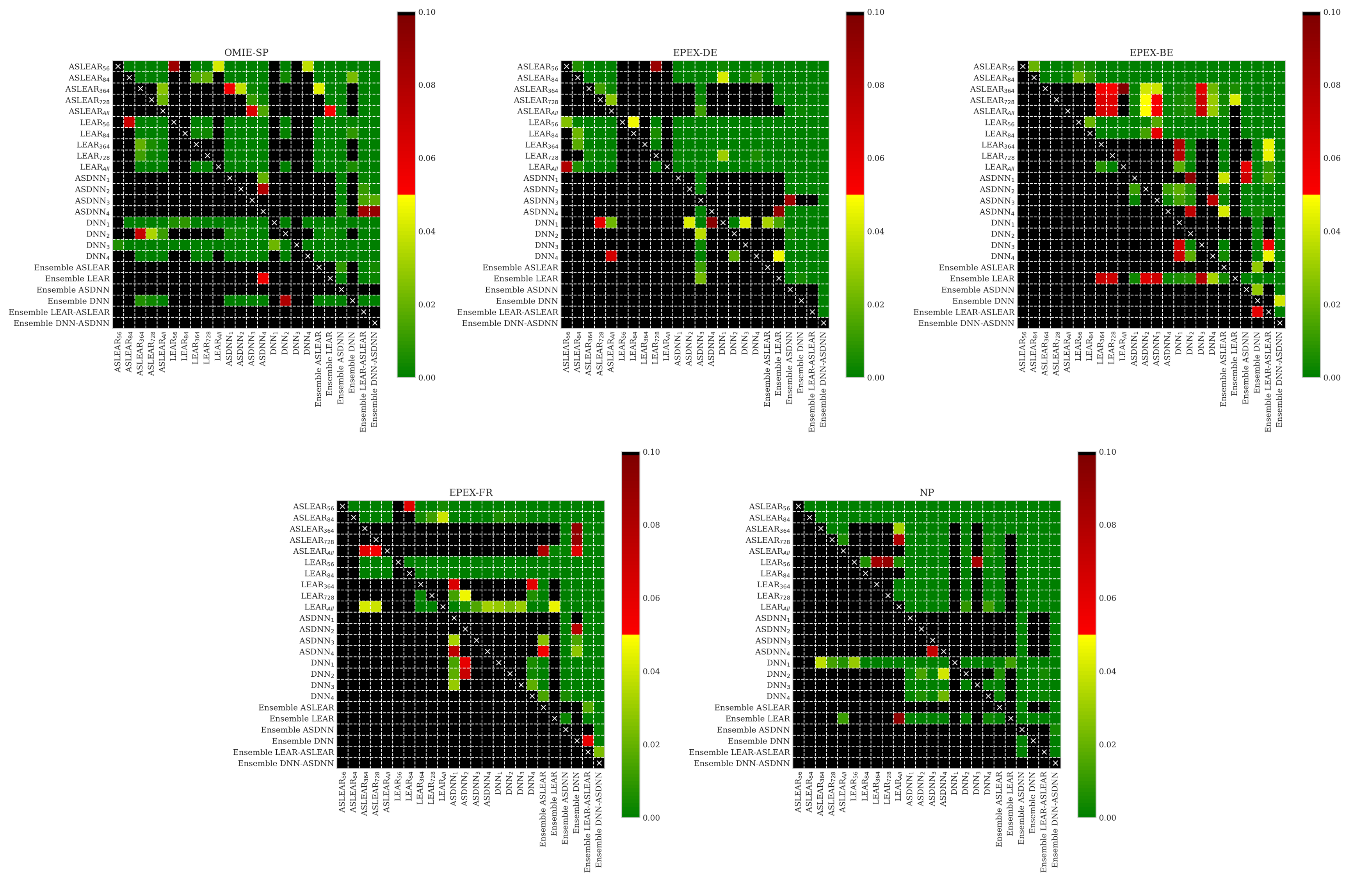}
\caption{Multivariate DM test between each analyzed model, including ensembles, for the every dataset considered}
\label{fig:dm}
\end{figure}

A statistical analysis is conducted for each market to compare the predictive performance of different models using the multidimensional Diebold-Mariano test. The results of this analysis are presented in the form of a colored matrix heatmap, where each cell represents a p-value. It is important to note that rejecting the null hypothesis indicates that the model in the column of the matrix performs better than the model in the corresponding row. Figure \ref{fig:dm} shows the results.\\

The statistical test confirms the results discussed in the previous section. The individual non-adaptive models do not perform particularly well compared to the other models. However, the ASDNN type models show very good performance generally, even outperforming ensembles in some situations. LEAR and ASLEAR models with small calibration windows (56 and 84) do not produce good results on their own and, in case of using a single model of this type larger windows should be considered. The Belgian market is the only one that does not seem to benefit from the use of adaptive versions, although it has been analysed in Figure 5 that this is probably related to an improved treatment of outliers. Ensembles significantly improve performance in general and the combination of DNN and ASDNN models stands out in all markets.

\section{Conclusions}\label{sec:conclusions}
In this work, a new framework for a price prediction model in the Day-Ahead market based on the price dynamics has been proposed. This new approach has been thoroughly studied, demonstrating improved results across various metrics and showing statistical improvement in five different markets and two distinct market periods. The outlier mitigation process has proven to be vital for achieving these results for the proposed models, although it could still be improved for better results. Additionally, two new and recent datasets have been made available to the community, aiming to explore new models on datasets that are closer to the current market situation. To get the best potential out of the models, combining them has proven to be key. In particular, combining the two methodologies evaluated produces the best results.\\

While the results may be further improved by using alternative learning algorithms for $f$ in Equation (\ref{eq:model}), in some cases no statistically significant difference has been observed among the evaluated individual adaptive models in this study, which are the ones that have performed the best by themselves. One way to achieve better results could be through the use of additional explanatory variables in estimating $\mu_d$ and $\sigma_d$. The inclusion of new explanatory features, especially those related to variable generation costs such as gas prices, oil prices, etc., can lead to improved results \citep{ortiz2016price, marcjasz2022distributional, shiri2015electricity}. In this context where frequent changes in data distribution can occur, the selection of variables must be done carefully \citep{sebastian2023feature}. It is crucial to consider the potential impact of dataset shifts and the biases that can be generated in this context by different explanatory variables.  Additionally, adaptive standardisation has been used, but other adaptive transformations could also be considered, like adaptive median-arcsinh. Finally, it has already been indicated that the combination of models produces very good results. This combination has been done by simply using the mean, which usually produces results that generalise very well \citep{petropoulos2022forecasting}. However, following the study in Figure \ref{fig:maes_by_month}, it can be seen that in some cases the best model alternates and is not always the same. This is why we think that online aggregation schemes such as those presented in \cite{gaillard2014second}, \cite{wintenberger2017optimal} or \cite{adjakossa2023kalman} should be studied in the context of EPF.

\section*{Declaration of competing interest}
The authors declare that they have no known competing financial interests or personal relationships that could have appeared to
influence the work reported in this paper.

\section*{CRediT authorship contribution statement}
\textbf{Carlos Sebastián}: Writing – original draft, Visualization, Software, Methodology, Conceptualization. \textbf{Carlos E. González-Guillén}: Writing – review \& editing, Validation, Supervision, Conceptualization, Project administration, Funding acquisition. \textbf{Jesús Juan}: Writing – review \& editing, Validation, Supervision, Conceptualization, Project administration, Funding acquisition.

\section*{Data availability}
The authors have shared the link to the repository, where the data and the code to reproduce the results are available. 

\section*{Funding}
This work has been funded by grant MIG-20211033 from Centro para el Desarrollo Tecnológico Industrial, Ministerio de Universidades, and European Union-NextGenerationEU. C.E.G.G. was also funded by a Re-qualification grant of Universidad Politécnica de Madrid funded by European Union-NextGenerationEU and by Ministerio de Universidades.

\newpage

\bibliographystyle{elsarticle-harv}

\newpage
\appendix

\section{Results of the LEAR model with LARS-AIC method for hyperparameter tuning}\label{ap:lars}

As the use of the LARS-AIC method has been replaced by cross-validation to calibrate the hyperparameter associated with regularisation in the LEAR method, Table \ref{table:LARS_AIC} shows the results comparing runs between methodologies to show that they are practically equivalent. To directly compare the results with those of the proposed methodology, the results are presented as a function of the ratio $\frac{\text{Metric using LARS-AIC}}{\text{Metric using CV}}$. Thus, if the result is greater than 1, it means that using the LARS-AIC procedure is worse than the CV strategy (and by what percentage) and if it is less than 1, it means that it is better (and by what percentage). We will call this ratio as the performance ratio.\\

\begin{table}[h]
\centering
\begin{tabular}{c|c|ccc|cc}
\multirow{2}{*}{\textbf{Market}} & \multirow{2}{*}{\textbf{Metrics}} & \multicolumn{3}{c|}{\textbf{ASLEAR}} & \multicolumn{2}{c}{\textbf{LEAR}} \\ \cline{3-7} 
 &  & \textbf{\begin{tabular}[c]{@{}c@{}}1092/\\ 364\end{tabular}} & \textbf{\begin{tabular}[c]{@{}c@{}}1456/\\ 728\end{tabular}} & \textbf{All} & \textbf{\begin{tabular}[c]{@{}c@{}}1092/\\ 364\end{tabular}} & \textbf{\begin{tabular}[c]{@{}c@{}}1456/\\ 728\end{tabular}} \\ \hline
\multirow{4}{*}{\textbf{\begin{tabular}[c]{@{}c@{}}OMIE\\ SP\end{tabular}}} & \textbf{MAE} & 1,0018 & 0,9941 & 1,0080 & 1,0770 & 1,0695 \\
 & \textbf{RMSE} & 1,0049 & 1,0012 & 1,0068 & 1,0651 & 1,0530 \\
 & \textbf{sMAPE} & 0,9998 & 0,9873 & 1,0260 & 1,0644 & 1,0479 \\
 & \textbf{rMAE} & 1,0047 & 0,9947 & 1,0059 & 1,0859 & 1,0627 \\ \hline
\multirow{4}{*}{\textbf{\begin{tabular}[c]{@{}c@{}}EPEX\\ DE\end{tabular}}} & \textbf{MAE} & 0,9713 & 0,9642 & 0,9692 & 1,0360 & 1,0628 \\
 & \textbf{RMSE} & 0,9787 & 0,9698 & 0,9675 & 1,0296 & 1,0649 \\
 & \textbf{sMAPE} & 1,0009 & 0,9728 & 0,9844 & 1,0244 & 1,0478 \\
 & \textbf{rMAE} & 0,9794 & 0,9697 & 0,9827 & 1,0330 & 1,0598 \\ \hline
\multirow{4}{*}{\textbf{\begin{tabular}[c]{@{}c@{}}EPEX\\ BE\end{tabular}}} & \textbf{MAE} & 1,0110 & 1,0081 & 1,0150 & 0,9916 & 1,0106 \\
 & \textbf{RMSE} & 1,0167 & 1,0041 & 1,0093 & 0,9926 & 0,9973 \\
 & \textbf{sMAPE} & 1,0269 & 1,0315 & 1,0340 & 1,0247 & 1,0651 \\
 & \textbf{rMAE} & 1,0079 & 1,0093 & 1,0104 & 0,9917 & 1,0078 \\ \hline
\multirow{4}{*}{\textbf{\begin{tabular}[c]{@{}c@{}}EPEX\\ FR\end{tabular}}} & \textbf{MAE} & 1,0152 & 1,0134 & 1,0237 & 1,0137 & 1,0263 \\
 & \textbf{RMSE} & 1,0340 & 1,0243 & 1,0463 & 0,9966 & 0,9922 \\
 & \textbf{sMAPE} & 1,0356 & 1,0400 & 1,0393 & 1,0243 & 1,0506 \\
 & \textbf{rMAE} & 1,0077 & 1,0083 & 1,0322 & 1,0095 & 1,0284 \\ \hline
\multirow{4}{*}{\textbf{NP}} & \textbf{MAE} & 1,0071 & 1,0083 & 1,0153 & 1,0551 & 1,0441 \\
 & \textbf{RMSE} & 1,0112 & 1,0126 & 1,0121 & 1,0449 & 1,0354 \\
 & \textbf{sMAPE} & 1,0208 & 1,0289 & 1,0358 & 1,0759 & 1,0829 \\
 & \textbf{rMAE} & 1,0101 & 1,0005 & 1,0075 & 1,0537 & 1,0530
\end{tabular}
\caption{Performance ratio when using the LARS-AIC methodology for hyperparameter selection}
\label{table:LARS_AIC}
\end{table}

The results are in line with what is stated in \cite{lago2021forecasting}: the results can be improved using a strategy based on cross-validation, but the computational cost is much higher. Moreover, this difference is minimal with the LARS-AIC procedure. The larger differences are observed in the most current datasets and in the NP market. The non-adaptive methodology is considerably improved in the case of using cross-validation. In the case of models based on the adaptive standardisation methodology, the EPEX-DE dataset is the only case where large improvements are obtained after hyperparameter determination. Overall, the results are generally similar and the methodologies are comparable.\\

\section{Results of the adaptive standardisation models without filtering outliers}\label{ap:results_no_filter}

To show the importance of the outlier filtering process in the models to which adaptive standardisation is applied, Table \ref{table:filtering_outliers_ASLEAR} shows the results without this process for the ASLEAR model. The results are shown using the performance ratio $\frac{\text{Metric without filtering outliers}}{\text{Metric filtering outliers}}$ as in the previous appendix.\\

\begin{table}[h]
\centering
\begin{tabular}{c|c|lllll}
\textbf{Market} & \textbf{Metrics} & \multicolumn{1}{c}{\textbf{56}} & \multicolumn{1}{c}{\textbf{84}} & \multicolumn{1}{c}{\textbf{\begin{tabular}[c]{@{}c@{}}1092/\\ 364\end{tabular}}} & \multicolumn{1}{c}{\textbf{\begin{tabular}[c]{@{}c@{}}1456/\\ 728\end{tabular}}} & \multicolumn{1}{c}{\textbf{All}} \\ \hline
\multirow{4}{*}{\textbf{\begin{tabular}[c]{@{}c@{}}OMIE\\ SP\end{tabular}}} & \textbf{MAE} & \multicolumn{1}{c}{1,0000} & \multicolumn{1}{c}{1,0000} & \multicolumn{1}{c}{1,0018} & \multicolumn{1}{c}{1,0017} & \multicolumn{1}{c}{1,0008} \\
 & \textbf{RMSE} & \multicolumn{1}{c}{1,0000} & \multicolumn{1}{c}{1,0000} & \multicolumn{1}{c}{1,0022} & \multicolumn{1}{c}{1,0003} & \multicolumn{1}{c}{0,9999} \\
 & \textbf{sMAPE} & \multicolumn{1}{c}{1,0000} & \multicolumn{1}{c}{1,0000} & \multicolumn{1}{c}{0,9998} & \multicolumn{1}{c}{0,9990} & \multicolumn{1}{c}{0,9979} \\
 & \textbf{rMAE} & \multicolumn{1}{c}{1,0000} & \multicolumn{1}{c}{1,0000} & \multicolumn{1}{c}{1,0018} & \multicolumn{1}{c}{1,0017} & \multicolumn{1}{c}{1,0008} \\ \hline
\multirow{4}{*}{\textbf{\begin{tabular}[c]{@{}c@{}}EPEX\\ DE\end{tabular}}} & \textbf{MAE} & \multicolumn{1}{c}{1,0000} & \multicolumn{1}{c}{1,0000} & \multicolumn{1}{c}{1,0000} & \multicolumn{1}{c}{1,0007} & \multicolumn{1}{c}{1,0179} \\
 & \textbf{RMSE} & \multicolumn{1}{c}{1,0000} & \multicolumn{1}{c}{1,0000} & \multicolumn{1}{c}{1,0000} & \multicolumn{1}{c}{1,0001} & \multicolumn{1}{c}{1,0199} \\
 & \textbf{sMAPE} & \multicolumn{1}{c}{1,0000} & \multicolumn{1}{c}{1,0000} & \multicolumn{1}{c}{1,0000} & \multicolumn{1}{c}{1,0005} & \multicolumn{1}{c}{1,0082} \\
 & \textbf{rMAE} & \multicolumn{1}{c}{1,0000} & \multicolumn{1}{c}{1,0000} & \multicolumn{1}{c}{1,0000} & \multicolumn{1}{c}{1,0007} & \multicolumn{1}{c}{1,0179} \\ \hline
\multirow{4}{*}{\textbf{\begin{tabular}[c]{@{}c@{}}EPEX\\ BE\end{tabular}}} & \textbf{MAE} & 1,2946 & 1,3579 & 1,1175 & 1,1252 & 1,1279 \\
 & \textbf{RMSE} & 2,4396 & 2,5351 & 1,1505 & 1,1567 & 1,1152 \\
 & \textbf{sMAPE} & 1,0986 & 1,1136 & 1,0913 & 1,0965 & 1,1090 \\
 & \textbf{rMAE} & 1,2946 & 1,3579 & 1,1175 & 1,1252 & 1,1279 \\ \hline
\multirow{4}{*}{\textbf{\begin{tabular}[c]{@{}c@{}}EPEX\\ FR\end{tabular}}} & \textbf{MAE} & 1,2862 & 1,2535 & 1,0980 & 1,1107 & 1,1111 \\
 & \textbf{RMSE} & 4,5893 & 3,7559 & 1,2325 & 1,0691 & 1,0609 \\
 & \textbf{sMAPE} & 1,0497 & 1,0490 & 1,0599 & 1,0778 & 1,0802 \\
 & \textbf{rMAE} & 1,2862 & 1,2535 & 1,0980 & 1,1107 & 1,1111 \\ \hline
\multirow{4}{*}{\textbf{NP}} & \textbf{MAE} & 1,0679 & 1,0677 & 1,0615 & 1,0547 & 1,0522 \\
 & \textbf{RMSE} & 1,1259 & 1,2046 & 1,0561 & 1,0522 & 1,0544 \\
 & \textbf{sMAPE} & 1,0576 & 1,0532 & 1,0574 & 1,0483 & 1,0447 \\
 & \textbf{rMAE} & 1,0679 & 1,0677 & 1,0615 & 1,0547 & 1,0522
\end{tabular}
\caption{Performance ratio of not filtering outliers for the ASLEAR model}
\label{table:filtering_outliers_ASLEAR}
\end{table}

For the ASDNN the results are shown in Table \ref{table:filtering_outliers_ASDNN}. As the hyperparameter configurations are not directly comparable, the best network without filtering outliers is compared to the best network with filtering outliers (indicated by 1 in the table), the second best network without filtering outliers is compared to the second best network with filtering outliers (indicated by 2) and so on. This way the performance ratio can be maintained, which makes the two situations easy to compare.\\

\begin{table}[h]
\centering
\begin{tabular}{c|c|llll}
\textbf{Market} & \textbf{Metrics} & \multicolumn{1}{c}{\textbf{1}} & \multicolumn{1}{c}{\textbf{2}} & \multicolumn{1}{c}{\textbf{3}} & \multicolumn{1}{c}{\textbf{4}} \\ \hline
\multirow{4}{*}{\textbf{\begin{tabular}[c]{@{}c@{}}OMIE\\ SP\end{tabular}}} & \textbf{MAE} & 1,0185 & 1,0080 & 1,0068 & 1,0496 \\
 & \textbf{RMSE} & 1,0212 & 1,0144 & 0,9888 & 1,0536 \\
 & \textbf{sMAPE} & 0,9991 & 1,0024 & 1,0266 & 1,0101 \\
 & \textbf{rMAE} & 1,0185 & 1,0080 & 1,0068 & 1,0496 \\ \hline
\multirow{4}{*}{\textbf{\begin{tabular}[c]{@{}c@{}}EPEX\\ DE\end{tabular}}} & \textbf{MAE} & 1,0149 & 0,9614 & 0,9541 & 0,9797 \\
 & \textbf{RMSE} & 1,0100 & 0,9594 & 0,9480 & 1,0105 \\
 & \textbf{sMAPE} & 1,0241 & 1,0112 & 0,9926 & 1,0348 \\
 & \textbf{rMAE} & 1,0149 & 0,9614 & 0,9541 & 0,9797 \\ \hline
\multirow{4}{*}{\textbf{\begin{tabular}[c]{@{}c@{}}EPEX\\ BE\end{tabular}}} & \textbf{MAE} & 1,0476 & 1,0673 & 1,0453 & 1,0927 \\
 & \textbf{RMSE} & 1,0255 & 1,0452 & 1,0079 & 1,0389 \\
 & \textbf{sMAPE} & 1,0482 & 1,0802 & 1,0695 & 1,0912 \\
 & \textbf{rMAE} & 1,0476 & 1,0673 & 1,0453 & 1,0927 \\ \hline
\multirow{4}{*}{\textbf{\begin{tabular}[c]{@{}c@{}}EPEX\\ FR\end{tabular}}} & \textbf{MAE} & 1,0759 & 1,0815 & 1,1041 & 1,1293 \\
 & \textbf{RMSE} & 1,0264 & 1,0303 & 1,0990 & 1,1591 \\
 & \textbf{sMAPE} & 1,0605 & 1,0333 & 1,0296 & 1,0635 \\
 & \textbf{rMAE} & 1,0759 & 1,0815 & 1,1041 & 1,1293 \\ \hline
\multirow{4}{*}{\textbf{NP}} & \textbf{MAE} & 1,0322 & 1,0313 & 1,0539 & 1,1493 \\
 & \textbf{RMSE} & 1,0139 & 1,0550 & 1,0756 & 3,3125 \\
 & \textbf{sMAPE} & 1,0098 & 1,0189 & 1,0773 & 1,0439 \\
 & \textbf{rMAE} & 1,0322 & 1,0313 & 1,0539 & 1,1493
\end{tabular}
\caption{Performance ratio of not filtering outliers for the ASDNN model}
\label{table:filtering_outliers_ASDNN}
\end{table}

In the case of ASLEAR, it can be seen that the OMIE-SP and EPEX-DE datasets are practically unaffected as they have very few outliers. However, the rest of the datasets are affected, especially EPEX-DE and EPEX-FR, which show very significant outliers. It can also be seen that the metric that increases the most is the RMSE, which is the least robust to outliers, so the change in performance is mainly due to outliers. In this case, outlier filtering is a good measure.\\

The same conclusion can be drawn for the ASDNN, although in this case the measure is not positive for EPEX-DE. The best network is obtained by filtering outliers, but the rest of the networks in this market are not better when filtering is done. Except for this case, in the rest of the markets the neural network also performs better when such preprocessing is done, also improving the results by a considerable percentage.\\

\newpage

\section{Results of the median-arcsinh transformation when filtering outliers}\label{ap:results_filter}

To show that the improvements come from the adaptive standardisation process and not from the outlier filtering, the experiments have been reproduced on the filtered series but with the median-arcsinh standardisation. The results are shown using the performance ratio $\frac{\text{Metric without filtering outliers}}{\text{Metric filtering outliers}}$ as in the previous appendices for both the LEAR model (Table \ref{table:filtering_outliers_LEAR}) and the DNN model (Table \ref{table:filtering_outliers_DNN}).\\

\begin{table}[h]
\centering
\begin{tabular}{c|c|lllll}
\textbf{Market} & \textbf{Metrics} & \multicolumn{1}{c}{\textbf{56}} & \multicolumn{1}{c}{\textbf{84}} & \multicolumn{1}{c}{\textbf{\begin{tabular}[c]{@{}c@{}}1092/\\ 364\end{tabular}}} & \multicolumn{1}{c}{\textbf{\begin{tabular}[c]{@{}c@{}}1456/\\ 728\end{tabular}}} & \textbf{All} \\ \hline
\multirow{4}{*}{\textbf{\begin{tabular}[c]{@{}c@{}}OMIE\\ SP\end{tabular}}} & \textbf{MAE} & 1,0000 & 1,0000 & 0,9999 & 0,9993 & 0,9995 \\
 & \textbf{RMSE} & 1,0000 & 1,0000 & 0,9997 & 0,9998 & 0,9993 \\
 & \textbf{sMAPE} & 1,0000 & 1,0000 & 0,9998 & 0,9999 & 1,0000 \\
 & \textbf{rMAE} & 1,0000 & 1,0000 & 0,9999 & 0,9993 & 1,0000 \\ \hline
\multirow{4}{*}{\textbf{\begin{tabular}[c]{@{}c@{}}EPEX\\ DE\end{tabular}}} & \textbf{MAE} & 1,0000 & 1,0000 & 1,0000 & 1,0004 & 0,9981 \\
 & \textbf{RMSE} & 1,0000 & 1,0000 & 1,0000 & 1,0003 & 0,9969 \\
 & \textbf{sMAPE} & 1,0000 & 1,0000 & 1,0000 & 1,0003 & 1,0000 \\
 & \textbf{rMAE} & 1,0000 & 1,0000 & 1,0000 & 1,0004 & 1,0000 \\ \hline
\multirow{4}{*}{\textbf{\begin{tabular}[c]{@{}c@{}}EPEX\\ BE\end{tabular}}} & \textbf{MAE} & 1,0086 & 1,0063 & 0,9977 & 0,9977 & 0,9969 \\
 & \textbf{RMSE} & 1,0751 & 1,0651 & 1,0118 & 1,0127 & 1,0141 \\
 & \textbf{sMAPE} & 0,9985 & 0,9991 & 0,9915 & 0,9929 & 1,0000 \\
 & \textbf{rMAE} & 1,0086 & 1,0063 & 0,9977 & 0,9977 & 1,0000 \\ \hline
\multirow{4}{*}{\textbf{\begin{tabular}[c]{@{}c@{}}EPEX\\ FR\end{tabular}}} & \textbf{MAE} & 1,0044 & 1,0045 & 1,0057 & 1,0055 & 1,0069 \\
 & \textbf{RMSE} & 1,1561 & 1,0944 & 1,0424 & 1,0468 & 1,0566 \\
 & \textbf{sMAPE} & 0,9997 & 1,0006 & 0,9970 & 0,9973 & 1,0000 \\
 & \textbf{rMAE} & 1,0044 & 1,0045 & 1,0057 & 1,0055 & 1,0000 \\ \hline
\multirow{4}{*}{\textbf{NP}} & \textbf{MAE} & 1,0040 & 0,9992 & 0,9991 & 1,0012 & 0,9948 \\
 & \textbf{RMSE} & 1,0094 & 1,0131 & 1,0086 & 1,0132 & 1,0085 \\
 & \textbf{sMAPE} & 1,0016 & 0,9970 & 0,9941 & 1,0004 & 0,9091 \\
 & \textbf{rMAE} & 1,0040 & 0,9992 & 0,9991 & 1,0012 & 1,0000
\end{tabular}
\caption{Performance ratio of not filtering outliers for the LEAR model}
\label{table:filtering_outliers_LEAR}
\end{table}

\begin{table}[h]
\centering
\begin{tabular}{c|c|llll}
\textbf{Market} & \textbf{Metrics} & \multicolumn{1}{c}{\textbf{1}} & \multicolumn{1}{c}{\textbf{2}} & \multicolumn{1}{c}{\textbf{3}} & \multicolumn{1}{c}{\textbf{4}} \\ \hline
\multirow{4}{*}{\textbf{\begin{tabular}[c]{@{}c@{}}OMIE\\ SP\end{tabular}}} & \textbf{MAE} & 0,9926 & 0,9375 & 0,9481 & 0,9172 \\
 & \textbf{RMSE} & 0,9859 & 0,9076 & 0,9651 & 0,9172 \\
 & \textbf{sMAPE} & 0,9955 & 0,9782 & 0,9508 & 0,9364 \\
 & \textbf{rMAE} & 0,9926 & 0,9375 & 0,9481 & 0,9172 \\ \hline
\multirow{4}{*}{\textbf{\begin{tabular}[c]{@{}c@{}}EPEX\\ DE\end{tabular}}} & \textbf{MAE} & 1,0175 & 1,0399 & 1,1356 & 1,1860 \\
 & \textbf{RMSE} & 1,0011 & 1,0265 & 1,2452 & 1,2203 \\
 & \textbf{sMAPE} & 1,0192 & 1,0038 & 1,1069 & 1,0718 \\
 & \textbf{rMAE} & 1,0175 & 1,0399 & 1,1356 & 1,1860 \\ \hline
\multirow{4}{*}{\textbf{\begin{tabular}[c]{@{}c@{}}EPEX\\ BE\end{tabular}}} & \textbf{MAE} & 1,0102 & 1,0132 & 0,9963 & 1,0494 \\
 & \textbf{RMSE} & 0,9722 & 1,0257 & 0,9828 & 1,0196 \\
 & \textbf{sMAPE} & 1,0263 & 1,0368 & 1,0115 & 1,0683 \\
 & \textbf{rMAE} & 1,0102 & 1,0132 & 0,9963 & 1,0494 \\ \hline
\multirow{4}{*}{\textbf{\begin{tabular}[c]{@{}c@{}}EPEX\\ FR\end{tabular}}} & \textbf{MAE} & 0,9981 & 0,9924 & 1,0084 & 1,0214 \\
 & \textbf{RMSE} & 0,9842 & 0,9876 & 1,0008 & 0,9978 \\
 & \textbf{sMAPE} & 1,0012 & 1,0220 & 1,0202 & 1,0209 \\
 & \textbf{rMAE} & 0,9981 & 0,9924 & 1,0084 & 1,0214 \\ \hline
\multirow{4}{*}{\textbf{NP}} & \textbf{MAE} & 0,9665 & 0,9870 & 0,9769 & 1,0457 \\
 & \textbf{RMSE} & 1,0042 & 1,0102 & 0,9823 & 1,0089 \\
 & \textbf{sMAPE} & 0,9577 & 0,9976 & 0,9749 & 1,0339 \\
 & \textbf{rMAE} & 0,9665 & 0,9870 & 0,9769 & 1,0457
\end{tabular}
\caption{Performance ratio of not filtering outliers for the DNN model}
\label{table:filtering_outliers_DNN}
\end{table}

For the LEAR model there is no noticeable difference between filtering or not filtering outliers using a ``traditional'' standardisation. For the DNN model there really are no major differences either. The EPEX-BE, EPEX-FR and NP datasets show slight differences, while for OMIE-SP and EPEX-DE they seem somewhat larger. The best networks in these two datasets behave equivalently and the differences are most noticeable in the worst hyperparameter settings. However, this could all be due to the randomness of neural networks, since in these two datasets the filtering of outliers is minimal compared to the others where the differences are negligible.\\

\end{document}